\documentstyle[12pt,epsfig]{article}
\begin{document}
\baselineskip=24pt
\def\rd{{\rm d}}
\newcommand{\Lamb}{\Lambda_{b}}
\newcommand{\Lamc}{\Lambda_{c}}
\newcommand{\omeb}{\omega_{b}^{(*)}}
\newcommand{\omec}{\omega_{c}^{(*)}}
\newcommand{\omeq}{\omega_{Q}^{(*)}}
\newcommand{\dsp}{\displaystyle}
\newcommand{\nn}{\nonumber}
\newcommand{\dfr}[2]{ \displaystyle\frac{#1}{#2} }
\newcommand{\Lag}{\Lambda \scriptscriptstyle _{ \rm GR} } 
\newcommand{\pa}{p\parallel}
\newcommand{\pe}{p\perp}
\newcommand{\pet}{p\top}
\newcommand{\paa}{p'\parallel}
\newcommand{\pee}{p'\perp} 
\newcommand{\pete}{p'\top} 
\renewcommand{\baselinestretch}{1.5}
\begin{titlepage}
\vspace{-20ex}
\vspace{1cm}
\begin{flushright}
\vspace{-3.0ex} 
    {\sf ADP-01-22/T456} \\
\vspace{5.0ex}
\end{flushright}

\centerline{\Large\sf Heavy Quark Distribution Functions in Heavy Baryons}
\vspace{6.4ex}
\centerline{\large\sf X.-H. Guo, A.W. Thomas and A.G. Williams}
\vspace{3.5ex}
\centerline{\sf Department of Physics and Mathematical Physics,}
\centerline{\sf and Special Research Center for the Subatomic Structure of
Matter,}
\centerline{\sf Adelaide University, SA 5005, Australia}
\centerline{\sf e-mail:  xhguo@physics.adelaide.edu.au,
athomas@physics.adelaide.edu.au,}
\centerline{\sf awilliam@physics.adelaide.edu.au}
\vspace{6ex}
\begin{center}
\begin{minipage}{5in}
\centerline{\large\sf 	Abstract}
\vspace{1.5ex}
\small {Using the Bethe-Salpeter (B-S) equations for heavy baryons
$\Lambda_Q$, $\Sigma_Q$, $\Xi_Q$ and $\Omega_Q$ ($Q=b$ or $c$), which were
established in previous work, we calculate the heavy quark distribution
functions in these baryons. The numerical results indicate that these 
distribution functions have an obvious peak at some fraction, $\alpha_0$, 
of the baryon's light-cone ``plus'' momentum component
carried by the heavy quark, and that as $m_Q$ becomes heavier
this peak becomes sharper and closer to 1. The dependence of the 
distribution functions on 
various input parameters in the B-S model is also discussed. The results are
seen to be qualitatively similar to an existing phenomenological model.}

\end{minipage}
\end{center}

\vspace{1cm}

\end{titlepage}
\vspace{0.2in}
{\large\bf I. Introduction}
\vspace{0.2in}

A significant amount of experimental data have been accumulated
on lepton nucleon deep inelastic scattering processes. 
From these data one can extract information on the 
parton distribution functions in the nucleon, which describe its 
nonperturbative hadronic structure. 
In comparison with the nucleon, much less is known
about the parton distribution functions in other baryons such as $\Lambda$,
$\Lambda_c$, and $\Lambda_b$. This is because
it is impossible to produce targets of these short lived
baryons suitable for experiments. Although the parton distribution functions in
these baryons cannot be studied through deep inelastic scattering processes,
it is still possible to obtain information on their parton distribution 
functions by measuring the fragmentation of quarks to baryons or the
decays of these baryons. In fact, the parton distribution functions in 
$\Lambda$, $\Sigma$ and $\Delta$ have been calculated in the MIT bag model 
dressed by mesons \cite{boros}. In this paper, we will study the heavy
quark distribution functions in heavy baryons.

The dynamics inside a heavy hadron is simplified by the fact that the light 
degrees of freedom in a heavy hadron are blind to the flavor and spin quantum
numbers of the heavy quark when its mass is much bigger than
the QCD scale \cite{wise}. This makes heavy flavor physics a good area for 
studying
nonperturbative QCD interactions. In fact, with more measurements
on heavy baryons becoming available \cite{opal,ua1,cdf2,lep}, theoretical 
study of the structure of heavy baryons is becoming increasingly important. 

The parton distribution functions are scale-dependent, and their  
evolution in the perturbative region is described by the famous Dokshitzer-
Gribov-Lipatov-Altarelli-Parisi (DGLAP) equations \cite{dglap}. 
However, to determine the parton distribution functions at some low energy 
scale (the boundary condition for DGLAP), one needs to
apply either lattice QCD \cite{detmold} or nonperturbative effective models. 
In the heavy quark limit,
the light degrees of freedom in a heavy baryon have good quantum numbers
which can be used to classify heavy baryons. Based on this fact,
in a previous work we took the heavy baryon to be composed of 
a heavy quark and a light diquark. With this picture, we established the B-S 
equations for the heavy baryons $\Lambda_Q$ and $\omega_{Q}$ (where
$\omega$ represents
$\Sigma$, $\Xi$, or $\Omega$), and solved these equations
numerically by assuming that their kernel contains a scalar confinement term 
and a one-gluon-exchange term \cite{bsguo}. It is the purpose of the present 
paper to calculate the heavy quark distribution functions in these heavy
baryons in our B-S formalism.

In $\Lambda_Q$ and $\omega_Q$ we have $0^+$ and $1^+$ diquarks, respectively.
Since in the heavy quark limit the internal 
dynamics of a heavy baryon are described
by the light degrees of freedom, we expect that the heavy quark 
distribution functions in $\omega_Q$ and $\omega^{*}_{Q}$ should be the same.
Furthermore, the differences among the heavy quark
distribution functions in $\Sigma_Q$, $\Xi_Q$, and $\Omega_Q$ should be
caused by $SU(3)$ flavor breaking effects. 

The remainder of this paper is organized as follows. In Section II we
derive the formulas for the heavy quark distribution functions in various
heavy baryons in the B-S formalism. In Section III we present numerical 
results for these distribution functions and discuss their dependence
on the parameters of the model. We also compare our results with the
distribution functions of Guo and Kroll \cite{guo}. Finally, we give a 
summary and some suggestions for future work in Section IV.

\vspace{0.2in}
{\large\bf II. Formalism for the heavy quark distribution functions in 
$\Lambda_Q$, $\Sigma_Q$, $\Xi_Q$ and $\Omega_Q$}
\vspace{0.2in}

Based on the picture that a heavy baryon is composed of a heavy quark and a 
light diquark, it was shown that the B-S equation for $\Lambda_Q$ is
\cite{bsguo}
\begin{equation}
\chi_P(p)=S_F(\lambda_1 P+p)\int \frac{\rd^4q}{(2\pi)^4}G(P,p,q)\chi_P(q)
S_D(-\lambda_2 P+p),
\label{2a}
\end{equation}
where $\chi_P(p)$ is the B-S wave function in momentum space,
$G(P,p,q)$ is the kernel, $S_F$ and $S_D$
are the propagators of the heavy quark and light scalar diquark, respectively,
with $\lambda_1=\frac{m_Q}{m_Q+m_D}$ and $\lambda_2=\frac{m_D}{m_Q+m_D}$
($m_Q$ and $m_D$ are the masses of the heavy quark and the light diquark,
respectively), $P$ is the momentum of $\Lambda_Q$,
and $p$ is the relative momentum of the two constituents.

Similarly, the B-S equation for $\omega_Q$ takes the form
\begin{equation}
\chi_{P}^{\mu}(p)=S_F(\lambda_1 P+p)\int \frac{\rd^4q}{(2\pi)^4}
G_{\rho\nu}(P,p,q)\chi_{P}^{\nu}(q)S_{D}^{\mu\rho}(-\lambda_2 P+p),
\label{2b}
\end{equation}
where $G_{\rho\nu}(P,p,q)$ is the kernel and $S_{D}^{\mu\rho}$
is the propagator of the $1^+$ diquark.

In the heavy quark limit
\begin{equation}
\chi_{0 P}(p)=\phi_{0 P}(p)u_{\Lambda_Q}(v,s),
\label{2c}
\end{equation}
where $\phi_{0 P}(p)$ is a scalar function, and $u_{\Lambda_Q}(v,s)$
is the Dirac spinor for $\Lambda_Q$ with helicity $s$ and velocity $v$.

For $\omega_Q$ there are three scalar functions, $A, C$ and $D$, in the 
B-S wave function
\begin{equation}
\chi_{P}^{\mu}=A B^{\mu}(v)+C v^{\mu}p_{t\nu} B^{\nu}(v)+
D p_{t}^{\mu}p_{t\nu} B^{\nu}(v),
\label{2d}
\end{equation}
where $B_{\mu}(v)=\frac{1}{\sqrt{3}}(\gamma_\mu +v_\mu)\gamma_5 u(v)$,
and $p_t\equiv p-(v\cdot p)v$. $B_{\mu}(v)$ satisfies the constraints
$\rlap /v B_{\mu}(v)=B_{\mu}(v)$ and $v^\mu B_{\mu}(v)=0$.

The B-S equations have been solved numerically in the covariant instantaneous 
approximation, assuming the kernels contain a scalar confinement term, 
$\tilde{V}_1$, and a one-gluon-exchange term, $\tilde{V}_2$, with the
following form
\begin{eqnarray}
\tilde{V}_1&=&\frac{8\pi\kappa}{[(p_t-q_t)^2+\mu^2]^2}-(2\pi)^3
\delta^3  (p_t-q_t)
	\int \frac{\rd^3 k}{(2\pi)^3}\frac{8\pi\kappa}{(k^2+\mu^2)^2}, \nn \\
\tilde{V}_2&=&-\frac{16\pi}{3}
	\frac{\alpha_{s}
^{({\rm eff}) 2}Q_{0}^{2}}{[(p_t-q_t)^2+\mu^2][(p_t-q_t)^2+Q_{0}^{2}]},
\label{2e}
\end{eqnarray}
where $\kappa$ and $\alpha_{s}^{({\rm eff})}$ are coupling parameters related
to scalar confinement and one-gluon-exchange,
respectively, where $Q_{0}^{2}$ is a parameter associated with the 
gluon-diquark
vertex, and where the parameter $\mu$ is introduced to avoid the infra-red 
divergence in numerical calculations, with the limit $\mu \rightarrow 0$ 
being taken at the end of the calculation.

The twist-2 heavy quark distribution function in $A^+=0$ gauge
is defined as \cite{collins, jaffe} 
\begin{equation}
Q(\alpha)=\sqrt{2}P^+ \int \frac{{\rm d}x^-}{2 \pi} e^{-i \alpha P^+ x^-}
\langle B | T \bar{\psi}_Q (x^-) \gamma^+ \psi_Q(0) | B \rangle,
\label{2f}
\end{equation}
where $\psi_Q$ is the field operator of the heavy quark $Q$, $P^+ = 
\frac{1}{\sqrt{2}}(P^0 +P^3)$,
$\gamma^+ = \frac{1}{\sqrt{2}}(\gamma^0 +\gamma^3)$, $\psi_Q(x^-)$
denotes $\psi_Q(x)$ at $x^+={\bf x}_{\perp}=0$, and $| B \rangle$ 
represents the heavy baryon state with the normalization 
$\langle B, {\bf P},\lambda | B , {\bf P'},\lambda^{\prime}\rangle
=(2\pi)^3 P_0/m_B \delta_{\lambda,\lambda^{\prime}} \delta^3 ({\bf P}
-{\bf P'})$ (we have chosen the normalization convention
$\bar{u}_{B} u_{B}=1$).
 
We define the two-point function 
\begin{equation}
M_{\beta \alpha} (P, k)= \int {\rm d}x^4 e^{-i k x}
\langle B | T \bar{\psi}_{Q \alpha} (x) \psi_{Q \beta}(0) | B \rangle
\label{2g}
\end{equation}
and then the heavy quark distribution function can be  expressed as
\begin{equation}
Q(\alpha)= \int \frac{{\rm d}^4 k}{(2\pi)^4} \sqrt{2}
P^+ \delta(k^+ - \alpha P^+) {\rm Tr}[\gamma^+ M(P, k)].
\label{2h}
\end{equation}

The parameter $\alpha$ in Eqs. (\ref{2f},\ref{2h}) corresponds to the fraction 
of the heavy baryon's light-cone momentum component, $P^+$, carried by the heavy
quark, $Q$. When it is in the range $0 \leq \alpha \leq 1$, $Q(\alpha)$
measures the probability to find the heavy quark with the ``plus'' momentum
fraction $\alpha$. In principle, in a heavy baryon there is a possibility
to find a heavy antiquark, which is generated from the QCD vacuum. However,
since we are considering heavy quarks with masses much larger than the
QCD scale $\Lambda_{\rm QCD}$, it is very difficult to produce them
from the QCD vacuum. Therefore, we
neglect the heavy antiquark distribution functions. In other words,  
the valence heavy quark distribution function is the same as
the heavy quark distribution function, $Q(\alpha)$. Since there is only
one heavy quark in a heavy baryon we have the following normalization condition
for $Q(\alpha)$:
\begin{equation}
\int {\rm d}\alpha Q(\alpha)=1.
\label{2h1}
\end{equation}

The two-point function $M(P, k)$ can be evaluated in our B-S framework.
After some algebra we obtain 
for $\Lambda_Q$ in the heavy quark limit the result
\begin{equation}
M_{\beta \alpha}^{\Lambda_Q} (P, k)=(\bar{u}_{\Lambda_Q})_{\alpha} 
[\phi_{0 P}(k-\lambda_1 P)]^2 S_{D}^{-1}(P-k) (u_{\Lambda_Q})_{\beta}.
\label{2i}
\end{equation}

The propagator of the light scalar diquark has the form
\begin{equation}
S_{D}=\frac{i}{p_l^2-W_p^2+i\epsilon},
\label{2j}
\end{equation}
where $p_l\equiv v\cdot p-\lambda_2 m_{\Lambda_Q}$ and 
$W_{p}\equiv \sqrt{p_{t}^{2}+m_{D}^{2}}$.

The propagator of the heavy quark in the heavy quark limit has the form
\begin{equation}
S_{F}=\frac{i (1 +\rlap /v)}{2(p_l+E_0+m_D+i\epsilon)},
\label{2jj}
\end{equation}
where $E_0$ is the binding energy in the heavy quark limit.

We choose to work in the rest frame of $\Lambda_Q$, in which we have
$k^0+k^3=\alpha m_{\Lambda_Q}$ ($m_{\Lambda_Q}$ is the mass of
$\Lambda_Q$) from the $\delta(k^+ - \alpha P^+)$ constraint
in Eq. (\ref{2h}). Hence we find
\begin{equation}
S_{D}^{-1}=-i[-|{\bf k}_{\perp}|^2+2(1-\alpha)m_{\Lambda_Q} k^3
+(1-\alpha)^2 m_{\Lambda_Q}^2-m_{D}^{2}].
\label{2k}
\end{equation}
The B-S wave function, $\phi_{0 P}(p)$, can be expressed in terms of 
$\tilde{\phi}_{0 P}(q_t)$ $\equiv \int \frac{{\rm d}p_l}{2\pi}\phi_{0 P}(p)$
as
\cite{bsguo}:
\begin{equation}
\phi_{0 P}(p)=\frac{i}{(p_l+E_0+m_D+i\epsilon)(p_{l}^{2}-W_{p}^{2}+i\epsilon)} 
\int \frac{\rd^3q_t}{(2\pi)^3}(\tilde{V}_1 +2p_l \tilde{V}_2)
\tilde{\phi}_{0 P}(q_t).
\label{2l}
\end{equation}
The constraint $\delta(k^+ - \alpha P^+)$ leads to $p_l=-k^3
+(\alpha-1)m_{\Lambda_Q}$, and $|p_t|^2=|{\bf k}|^2$.
Substituting Eqs. (\ref{2i},\ref{2l}) into Eq. (\ref{2h}), 
we integrate out  $k^0$ 
with the aid of $\delta(k^+ - \alpha P^+)$. Furthermore, the component 
$k^3$ can also
be integrated out by choosing the appropriate contour. 
Then we arrive at the following result:
\begin{eqnarray}
Q^{\Lambda_Q}(\alpha)&=&\frac{1}{2\sqrt{2} \pi (1-\alpha)} 
\int \frac{{\rm d}^2 {\bf k}_\perp}{(2\pi)^2}\frac{1}{[E_0+m_D+\frac{1}{2}
(\alpha-1)m_{\Lambda_Q}+\frac{1}{2(\alpha-1)m_{\Lambda_Q}}(|{\bf k}_{\perp}|^2
+m_D^2)]^2} \nn\\
&&\left\{\int \frac{{\rm d}^3 q_t}{(2\pi)^3}\left[\tilde{V}_1(p_t-q_t)
+\left((\alpha-1)m_{\Lambda_Q}+\frac{1}{(\alpha-1)m_{\Lambda_Q}}
(|{\bf k}_{\perp}|^2 +m_D^2)\right) \right.\right.\nn\\
&&\left.\left.\tilde{V}_2(p_t-q_t)\right]
\tilde{\phi}_{0 P}(q_t)\right\}^2,
\label{2m}
\end{eqnarray}
where $|p_t|^2=|{\bf k}_{\perp}|^2+(k^{3}_{\rm pole})^2$ and where 
$k^{3}_{\rm pole}=\frac{1}{2(\alpha-1)m_{\Lambda_Q}}
[(\alpha-1)^2m_{\Lambda_Q}^2-|{\bf k}_{\perp}|^2-m_D^2]$.

Substituting $\tilde{V}_1$ and $\tilde{V}_2$ in Eq. (\ref{2e}) into 
Eq. (\ref{2m}) and integrating out the angular coordinates we finally obtain
\begin{eqnarray}
Q^{\Lambda_Q}(\alpha)&=&\frac{1}{2\sqrt{2} \pi (1-\alpha)} 
\int \frac{|{\bf k}_{\perp}|{\rm d}|{\bf k}_{\perp}|}{2\pi}\frac{1}
{[E_0+m_D+\frac{1}{2}
(\alpha-1)m_{\Lambda_Q}+\frac{1}{2(\alpha-1)m_{\Lambda_Q}}(|{\bf k}_{\perp}|^2
+m_D^2)]^2} \nn\\
&&\left(\int \frac{q_t^2{\rm d} q_t}{4\pi^2}\left\{\frac{16\pi \kappa}
{(p_t^2+q_t^2+\mu^2)^2-4p_t^2 q_t^2}[\tilde{\phi}_{0 P}(q_t)
-\tilde{\phi}_{0 P}(p_t)]\right.\right.\nn\\
&&\left.\left. +\frac{16\pi\alpha_{s}^{({\rm eff}) 2}Q_{0}^{2}}
{3(Q_{0}^{2}-\mu^2)}\left[(\alpha-1)m_{\Lambda_Q}+\frac{1}{(\alpha-1)
m_{\Lambda_Q}}(|{\bf k}_{\perp}|^2 +m_D^2)\right]\right.\right.\nn\\
&&\times \left.\left.
 \frac{1}{2|p_t||q_t|}\left[{\rm ln}\frac{(|p_t|+|q_t|)^2+\mu^2}
{(|p_t|-|q_t|)^2+\mu^2}-{\rm ln}\frac{(|p_t|+|q_t|)^2+Q_{0}^{2}}
{(|p_t|-|q_t|)^2+Q_{0}^{2}}\right]\tilde{\phi}_{0 P}(q_t)\right\}\right)^2,
\label{2n}
\end{eqnarray}
where 
\begin{equation}
|p_t|=\frac{1}{2(1-\alpha)m_{\Lambda_Q}}\sqrt{[(\alpha-1)^2m_{\Lambda_Q}^2
-|{\bf k}_{\perp}|^2-m_D^2]^2+4(\alpha-1)^2|{\bf k}_{\perp}|^2m_{\Lambda_Q}^2}.
\label{2o}
\end{equation}
In deriving Eq. (\ref{2n}) we have used the following equations to
reduce the three dimensional integrations 
to one dimensional integrations 
\begin{equation}
\int \frac{\rd^3 q_t}{(2\pi)^3}\frac{\rho(q_{t}^{2})}{[(p_t-q_t)^2
+\mu^2]^2}=\int \frac{q_{t}^{2}\rd q_t}{4\pi^2}
\frac{2\rho(q_{t}^{2})}{(p_{t}^{2}+q_{t}^{2}+\mu^2)^2
-4p_{t}^{2}q_{t}^{2}},
\label{2n1}
\end{equation}
and
\begin{equation}
\int \frac{\rd^3 q_t}{(2\pi)^3}\frac{\rho(q_{t}^{2})}{(p_t-q_t)^2
+\delta^2}=\int\frac{q_{t}^{2}\rd q_t}{4\pi^2}
\frac{\rho(q_{t}^{2})}{2|p_{t}||q_{t}|}{\rm ln}
\frac{(|p_{t}|+|q_{t}|)^{2}+\delta^2}{(|p_{t}|-|q_{t}|)^{2}+\delta^2},
\label{2n2}
\end{equation}
where $\rho(q_{t}^{2})$ is some arbitrary scalar function of $q_{t}^{2}$. 

Now we turn to $\omega_Q$. The two-point function $M(P, k)$ in this case
can be derived in a similar way and the result is
\begin{equation}
M_{\beta \alpha}^{\omega_Q} (P, k)=\bar{\chi}_{\alpha}^{\mu} 
(k-\lambda_1 P) S_{D \mu\nu}^{-1}(P-k) \chi_{\beta}^{\nu}(k-\lambda_1 P).
\label{2p}
\end{equation}

Substituting the B-S equation (\ref{2b}) into Eq. (\ref{2p}), using 
Eqs. (\ref{2d}) and (\ref{2jj}), and working in the covariant instantaneous
approximation, $p_l=q_l$ (which ensures that the B-S equation is still
covariant after this approximation), we have
\begin{eqnarray}
{\rm Tr} [\gamma^+ M^{\omega_Q} (P, k)]&=&\frac{1}{p_l+E_0+m_D+i\epsilon}
[A \bar{B}^{\mu}+C v^{\mu}p_{t}\cdot \bar{B}+D p_{t}^{\mu}p_{t}\cdot 
\bar{B}]\gamma^+ \nn\\
&&\int \frac{{\rm d}^3 q_t}{(2\pi)^3}\left\{B_{\mu} \left[\tilde{A}
(\tilde{V}_1+2p_l\tilde{V}_2)-\tilde{C}\frac{(p_t\cdot q_t)^2-p_{t}^{2}
q_{t}^{2}}{2p_{t}^{2}}\tilde{V}_2 \right. \right.\nn\\
&&\left.\left.+\tilde{D}\frac{(p_t\cdot q_t)^2-p_{t}^{2}
q_{t}^{2}}{2p_{t}^{2}}(\tilde{V}_1+2p_l\tilde{V}_2)\right]\right. \nn\\
&&+\left. v_\mu p_t \cdot B\left[-\tilde{A}\tilde{V}_2-\frac{p_t\cdot q_t}
{p_{t}^{2}}\tilde{C}\tilde{V}_1+\tilde{D}\frac{(p_t\cdot q_t)^2}{p_{t}^{2}}
\tilde{V}_2\right]\right.\nn\\
&&\left. +p_{t \mu} p_t \cdot B \frac{3(p_t\cdot q_t)^2-p_{t}^{2}q_{t}^{2}}
{2p_{t}^{4}}[-\tilde{C} \tilde{V}_2+\tilde{D}(\tilde{V}_1+2p_l\tilde{V}_2)]
\right\}.\nn\\
&&\label{2q}
\end{eqnarray}

As for $\Lambda_Q$, the B-S wave functions 
$A(p_l,p_{t}^{2})$, $C(p_l,p_{t}^{2})$ and $D(p_l,p_{t}^{2})$ are
related to $\tilde{A}(p_{t}^{2})$, $\tilde{C}(p_{t}^{2})$ and 
$\tilde{D}(p_{t}^{2})$ through the following equations \cite{bsguo}:
\begin{eqnarray}
A(p_l,p_{t}^{2})&=&\frac{-i}{(p_l+E_0+m_D+i\epsilon)(p_{l}^{2}-W_{p}^{2}
+i\epsilon)}\int \frac{\rd^3 q_t}
{(2\pi)^3}\left\{\tilde{A}(q_{t}^{2})(\tilde{V}_1+2p_l\tilde{V}_2)\right.\nn\\
&&\left.-\tilde{C}(q_{t}^{2})\frac{(p_t\cdot q_t)^2-p_{t}^{2}
q_{t}^{2}}{2p_{t}^{2}}\tilde{V}_2 
+\tilde{D}(q_{t}^{2})\frac{(p_t\cdot q_t)^2-p_{t}^{2}
q_{t}^{2}}{2p_{t}^{2}}(\tilde{V}_1+2p_l\tilde{V}_2)\right\},\nn\\
&& 
\label{2r1}
\end{eqnarray}
\begin{eqnarray}
C(p_l,p_{t}^{2})&=&\frac{-i}{m_{D}^{2}(p_l+E_0+m_D+i\epsilon)
(p_{l}^{2}-W_{p}^{2}+i\epsilon)}
\int \frac{\rd^3 q_t}
{(2\pi)^3} 
\left\{-\tilde{A}(q_{t}^{2})[p_l\tilde{V}_1 \right.\nn\\ 
&&\left.+(p_{l}^{2}+m_{D}^{2})\tilde{V}_2]
-\tilde{C}(q_{t}^{2})\left[(p_{l}^{2}-m_{D}^{2})
\frac{p_t\cdot q_t}{p_{t}^{2}}
\tilde{V}_1+p_l\frac{(p_t\cdot q_t)^2}{p_{t}^{2}} 
\tilde{V}_2\right]\right. \nn\\
&&\left.+\tilde{D}(q_{t}^{2})\frac{(p_t\cdot q_t)^2}{p_{t}^{2}}
[p_l\tilde{V}_1+(p_{l}^{2}+m_{D}^{2})
\tilde{V}_2]\right\},
\label{2r2}
\end{eqnarray}
\begin{eqnarray}
D(p_l,p_{t}^{2})&=&\frac{i}{m_{D}^{2}(p_l+E_0+m_D+i\epsilon)
(p_{l}^{2}-W_{p}^{2}+i\epsilon)}
\int \frac{\rd^3 q_t}
{(2\pi)^3}\left\{\tilde{A}(q_{t}^{2})(\tilde{V}_1+p_l\tilde{V}_2)\right.\nn\\
&&\left.+\tilde{C}(q_{t}^{2})\left[\frac{p_t\cdot q_t}{p_{t}^{2}}
p_l\tilde{V}_1+\frac{m_{D}^{2}(3(p_t\cdot q_t)^2-p_{t}^{2}
q_{t}^{2})+2p_{t}^{2}(p_t\cdot q_t)^2}{2p_{t}^{4}}
\tilde{V}_2\right]\right. \nn\\
&&\left.+\tilde{D}(q_{t}^{2})\left[-\frac{m_{D}^{2}(3(p_t\cdot q_t)^2-p_{t}^{2}
q_{t}^{2})+2p_{t}^{2}(p_t\cdot q_t)^2}{2p_{t}^{4}}
(\tilde{V}_1+2p_l\tilde{V}_2)\right.\right.\nn\\
&&\left.\left.+\frac{(p_t\cdot q_t)^2}{p_{t}^{2}}
p_l\tilde{V}_2\right]\right\},
\label{2r3}
\end{eqnarray}
and $\tilde{A}(p_{t}^{2})$, $\tilde{C}(p_{t}^{2})$ and 
$\tilde{D}(p_{t}^{2})$ obey the following
three coupled integral equations 
\begin{eqnarray}
\tilde{A}(p_{t}^{2})&=&\frac{-1}{2W_p(E_0+m_D-W_p)}\int \frac{\rd^3 q_t}
{(2\pi)^3}\left\{\tilde{A}(q_{t}^{2})(\tilde{V}_1-2W_p\tilde{V}_2)\right.\nn\\
&&\left.-\tilde{C}(q_{t}^{2})\frac{(p_t\cdot q_t)^2-p_{t}^{2}
q_{t}^{2}}{2p_{t}^{2}}\tilde{V}_2 
+\tilde{D}(q_{t}^{2})\frac{(p_t\cdot q_t)^2-p_{t}^{2}
q_{t}^{2}}{2p_{t}^{2}}(\tilde{V}_1-2W_p\tilde{V}_2)\right\}, \nn\\
&& 
\label{2s1}
\end{eqnarray}
\begin{eqnarray}
\tilde{C}(p_{t}^{2})&=&\frac{-1}{2m_{D}^{2}W_p(E_0+m_D-W_p)}
\int \frac{\rd^3 q_t}
{(2\pi)^3}\left\{\tilde{A}(q_{t}^{2})[W_p\tilde{V}_1 \right. \nn\\
&&\left.
-((E_0+m_D)W_p+m_{D}^{2})\tilde{V}_2]\right.\nn\\
&&\left.+\tilde{C}(q_{t}^{2})\left[-\frac{p_t\cdot q_t}{p_{t}^{2}}
((E_0+m_D)W_p-m_{D}^{2})\tilde{V}_1+W_p\frac{(p_t\cdot q_t)^2}{p_{t}^{2}} 
\tilde{V}_2\right]\right. \nn\\
&&\left.+\tilde{D}(q_{t}^{2})\frac{(p_t\cdot q_t)^2}{p_{t}^{2}}
[-W_p\tilde{V}_1+((E_0+m_D)W_p+m_{D}^{2})
\tilde{V}_2]\right\},
\label{2s2}
\end{eqnarray}
\begin{eqnarray}
\tilde{D}(p_{t}^{2})&=&\frac{1}{2m_{D}^{2}W_p(E_0+m_D-W_p)}
\int \frac{\rd^3 q_t}
{(2\pi)^3}\left\{\tilde{A}(q_{t}^{2})(\tilde{V}_1-W_p\tilde{V}_2)\right.\nn\\
&&\left.+\tilde{C}(q_{t}^{2})\left[-\frac{p_t\cdot q_t}{p_{t}^{2}}
W_p\tilde{V}_1+\frac{m_{D}^{2}(3(p_t\cdot q_t)^2-p_{t}^{2}
q_{t}^{2})+2p_{t}^{2}(p_t\cdot q_t)^2}{2p_{t}^{4}}
\tilde{V}_2\right]\right. \nn\\
&&\left.-\tilde{D}(q_{t}^{2})\left[\frac{m_{D}^{2}(3(p_t\cdot q_t)^2-p_{t}^{2}
q_{t}^{2})+2p_{t}^{2}(p_t\cdot q_t)^2}{2p_{t}^{4}}
(\tilde{V}_1-2W_p\tilde{V}_2)\right.\right.\nn\\
&&\left.\left.+\frac{(p_t\cdot q_t)^2}{p_{t}^{2}}
W_p\tilde{V}_2\right]\right\}.
\label{2s3}
\end{eqnarray}

Once again, we first integrate out $k^0$
with the help of the constraint $\delta(k^+ - \alpha P^+)$, then we further
integrate out $k^3$ by selecting the proper contour which contains 
the pole in $k^3$. With the aid of Eqs. (\ref{2r1}-\ref{2s3}), and noticing
that $\bar{B}^\mu \gamma^+ B_\mu=-\frac{1}{\sqrt{2}}$ and 
$p_t \cdot B \gamma^+ p_t \cdot B=\frac{1}{3\sqrt{2}}|{\bf k}|^2$, we
obtain: 
\begin{eqnarray}
Q^{\omega_Q}(\alpha)&=&\frac{1}{3\sqrt{2} \pi (1-\alpha)} 
\int \frac{|{\bf k}_{\perp}|{\rm d}|{\bf k}_{\perp}|}{2\pi}\frac{W_p}
{E_0+m_D-W_p} [-3 \tilde{A}(p_{t}^{2}) f_1(p_{t}^{2})\nn\\
&&+p_{t}^{2} \tilde{C}(p_{t}^{2}) f_2(p_{t}^{2})
+p_{t}^{2} \tilde{D}(p_{t}^{2}) f_3(p_{t}^{2})],
\label{2t}
\end{eqnarray}
where $|p_t|$ is given in Eq. (\ref{2o}), and 
\begin{equation}
f_1(p_{t}^{2})=\int \frac{\rd^3 q_t}
{(2\pi)^3}\left\{\tilde{A}(q_{t}^{2})(\tilde{V}_1-2W_p\tilde{V}_2)
-\frac{1}{3}q_{t}^{2}[-\tilde{C}(q_{t}^{2})\tilde{V}_2
+\tilde{D}(q_{t}^{2})(\tilde{V}_1-2W_p\tilde{V}_2)]\right\},
\label{2u1}
\end{equation}
\begin{equation}
f_2(p_{t}^{2})=\int \frac{\rd^3 q_t}
{(2\pi)^3}\left[-\tilde{A}(q_{t}^{2})\tilde{V}_2
+\tilde{C}(q_{t}^{2})\frac{p_t \cdot q_t}{p_{t}^{2}}\tilde{V}_1
+\tilde{D}(q_{t}^{2})\frac{(p_t \cdot q_t)^2}{p_{t}^{2}}\tilde{V}_2\right],
\label{2u2}
\end{equation}
\begin{equation}
f_3(p_{t}^{2})=\int \frac{\rd^3 q_t}
{(2\pi)^3}\left\{\tilde{A}(q_{t}^{2})(\tilde{V}_1-2W_p\tilde{V}_2)
-\frac{(p_t \cdot q_t)^2}{p_{t}^{2}}[-\tilde{C}(q_{t}^{2})\tilde{V}_2
+\tilde{D}(q_{t}^{2})(\tilde{V}_1-2W_p\tilde{V}_2)]\right\}.
\label{2u3}
\end{equation}

So far we have been working with the heavy quark limit, $m_Q \rightarrow
\infty$, of the B-S equation, but in Ref. \cite{bsguo} we also 
considered the $1/m_Q$ corrections
to the B-S equation for $\Lambda_Q$. To order $1/m_Q$, the heavy quark
distribution function in $\Lambda_Q$ is given by
$$Q^{\Lambda_Q}(\alpha)+\Delta Q^{\Lambda_Q}(\alpha)$$ 
where $ \Delta Q^{\Lambda_Q}(\alpha)$ denotes the $1/m_Q$ corrections. 
To order $1/m_Q$, two more scalar functions appear in the B-S wave function, 
which can be related to $\phi_{0 P}(p)$ in the
model of Ref. \cite{bsguo}. 
Consequently, the B-S wave function to order $1/m_Q$ is given by
\begin{equation}
\chi_{P}(p)=\phi_{0 P}(p)\left[1+\frac{\rlap / p_t}{2m_Q}\right]
u_{\Lambda_Q}(v,s).
\label{2v}
\end{equation}
 
With Eq. (\ref{2v}) we can evaluate the two-point function 
$M_{\beta \alpha}^{\Lambda_Q} (P, k)$ to order $1/m_Q$ from Eq. (\ref{2i}). 
Note
that the scalar diquark propagator in Eq. (\ref{2j}) remains unchanged to
order $1/m_Q$. From Eq. (\ref{2h}), after integrating out $k^0$ and $k^3$, 
we have
\begin{eqnarray}
\Delta Q^{\Lambda_Q}(\alpha)&=&\frac{1}{2\sqrt{2}  \pi (1-\alpha)} 
\int \frac{|{\bf k}_{\perp}|{\rm d}|{\bf k}_{\perp}|}{2\pi}
\frac{\frac{1}{2}
(\alpha-1)m_{\Lambda_Q}-\frac{1}{2(\alpha-1)m_{\Lambda_Q}}(|{\bf k}_{\perp}|^2
+m_D^2)}
{[E_0+m_D+\frac{1}{2}
(\alpha-1)m_{\Lambda_Q}+\frac{1}{2(\alpha-1)m_{\Lambda_Q}}(|{\bf k}_{\perp}|^2
+m_D^2)]^2} \nn\\
&&\left(\int \frac{q_t^2{\rm d} q_t}{4\pi^2}\left\{\frac{16\pi \kappa}
{(p_t^2+q_t^2+\mu^2)^2-4p_t^2 q_t^2}[\tilde{\phi}_{0 P}(q_t)
-\tilde{\phi}_{0 P}(p_t)]\right.\right.\nn\\
&&\left.\left. +\frac{16\pi\alpha_{s}^{({\rm eff}) 2}Q_{0}^{2}}
{3(Q_{0}^{2}-\mu^2)}\left[(\alpha-1)m_{\Lambda_Q}+\frac{1}{(\alpha-1)
m_{\Lambda_Q}}(|{\bf k}_{\perp}|^2 +m_D^2)\right]\right.\right.\nn\\
&&\times \left.\left.
 \frac{1}{2|p_t||q_t|}\left[{\rm ln}\frac{(|p_t|+|q_t|)^2+\mu^2}
{(|p_t|-|q_t|)^2+\mu^2}-{\rm ln}\frac{(|p_t|+|q_t|)^2+Q_{0}^{2}}
{(|p_t|-|q_t|)^2+Q_{0}^{2}}\right]\tilde{\phi}_{0 P}(q_t)\right\}\right)^2,
\nn\\
&&
\label{2w}
\end{eqnarray}
where again $|p_t|$ is given in Eq. (\ref{2o}).

The heavy quark distribution functions have been derived at some
scale $\nu_0$, which is of the order of $\Lambda_{\rm QCD}$.
The QCD running of these  functions is controlled by the DGLAP equations.
In our numerical calculations we will use the evolution
code provided in Ref. \cite{miya}
to give the heavy quark distribution functions at any higher scale.
This will be done to the next-to-leading order. Since the QCD interactions
are flavor independent, we can directly apply this code to the cases of
heavy quarks. It should be pointed out that the scale  $\nu_0$ cannot be
determined in our approach. We will treat it as a free parameter and
leave it to be determined by the future experimental data.

\vspace{0.2in}
{\large\bf III. Numerical results}
\vspace{0.2in}

In this section we will give numerical results for the heavy quark 
distribution functions based on the formulas presented in Section II. 
The B-S wave functions for $\Lambda_Q$ and $\omega_Q$ were solved numerically
in our previous work by discretizing the integration region (0, $\infty$) 
into $n$ pieces ($n$ is chosen to be sufficiently large and we use $n$-point 
Gauss quadrature rule to evaluate the integrals) and solving the 
eigenvalue equation
with the kernel $\tilde{V}_1$ and $\tilde{V}_2$ in Eq. (\ref{2e})
in the covariant instantaneous approximation \cite{bsguo}. The normalization
constants of these B-S wave functions are determined by the normalization
of Isgur-Wise functions at the zero-recoil point. Substituting 
these numerical solutions into Eqs. (\ref{2n},\ref{2t},\ref{2w}), we can 
obtain numerical results for the heavy quark distribution functions in 
$\Lambda_Q$ and $\omega_Q$.

In our model we have several parameters, i.e., $\alpha_{s}^{({\rm eff})}$, 
$\kappa$, 
$Q_{0}^{2}$, $m_D$ and $E_0$. The parameter $Q_{0}^{2}$ is chosen to be
$3.2$GeV$^2$ from the data for the electromagnetic form factor of the proton
\cite{bsguo,kroll}.  The parameters $\alpha_{s}^{({\rm eff})}$
and $\kappa$ are related to each other
when we solve the eigenvalue equations with fixed eigenvalues \cite{bsguo}. 
The parameter $\kappa$ is of the order $\Lambda_{\rm QCD}\kappa'$ where
$\kappa'$ is around 0.2GeV$^2$ from the potential model \cite{eichten,dai}. 
Thus we let $\kappa$ vary in the region between 0.02GeV$^3$ and 0.1GeV$^3$ 
\cite{bsguo}. The parameters $m_D$ and $E_0$ are constrained by the relation 
$m_D+E_0+\frac{1}{m_Q}E_1=m_{\Lambda_Q}-m_Q$ for $\Lambda_Q$
(to order $1/m_Q$) and $m_D+E_0=m_{\omega_Q}-m_Q$ for $\omega_Q$
($m_Q \rightarrow \infty$). From the heavy quark effective theory,  
$m_D+E_0$ and $E_1$ are independent of the flavor of the
heavy quark. It was shown \cite{bsguo}
that the value of $E_1$ (which is of the order
$\Lambda_{\rm QCD} E_0$) influences the numerical results only slightly.
In our numerical calculations we use $m_b=5.02$GeV and $m_c=1.58$GeV
which led to consistent predictions with experiments
from the B-S equation solutions in the meson case \cite{dai}. Consequently
we have $m_D+E_0+\frac{1}{m_b}E_1=0.62$GeV for $\Lambda_Q$ (where we have
neglected $1/m_b^2$ corrections), and $m_D+E_0=0.88$GeV for $\Sigma_Q$,
1.07GeV for $\Xi_{Q}$, 1.12GeV for $\Omega_{Q}$ in the heavy quark limit.
The parameter $m_D$ cannot be determined and hence we let it vary within some 
reasonable range. For $\Lambda_Q$, we choose $m_D$ to be in the range
0.65GeV $\sim$ 0.75GeV \cite{bsguo,tony}. The axial-vector diquark mass is 
chosen to vary from 0.9GeV to 1GeV for $\Sigma_{Q}$,
from 1.1GeV to 1.2GeV for $\Xi_{Q}$, and from 1.15GeV to 1.25GeV for 
$\Omega_{Q}$ \cite{bsguo}. With these choices for $m_D$, the binding energy
$E_0$ is negative and varies from around -30MeV to -130MeV.  

With these parameters, from Eqs. (\ref{2n},\ref{2t},\ref{2w}), we obtain
numerical results for the heavy quark distribution functions in $\Lambda_Q$
and $\omega_Q$. These results are shown in Figs.1 to 5. Figs.1(a),
2(a), 3(a), 4(a) and 5(a) show the dependence on $\kappa$ for a typical 
$m_D$, and the dependence on $m_D$ for $\kappa$=0.06GeV$^3$. All these
results correspond to some low energy scale which is of order several hundred
MeV. They are evolved to some higher scale using the DGLAP equations.
In Figs.1(b), 2(b), 3(b), 4(b) and 5(b), we show the heavy quark 
distribution functions at the scale $\nu^2=10$GeV$^2$, where we have taken
the low energy scale $\nu_0^2$ to be 0.25GeV$^2$, as an example. From these 
figures we can see that for different heavy baryons with the same heavy quark
flavor the shapes of the heavy quark distribution functions are rather 
similar. For the heavy quark distribution functions at the hadronic scale, 
there is an obvious peak at some ``plus'' 
momentum fraction carried by the heavy quark, $\alpha_0$, and 
this peak is much sharper for $b$-baryons than $c$-baryons.
Furthermore, $\alpha_0$ is much closer to 1 for $b$-baryons than $c$-baryons.
It can also be seen that QCD evolution makes the amplitudes of the peaks
much smaller. However, the distinction between $b$-quark and $c$-quark
distribution functions is still obvious at high $\nu^2$. From Figs. 1 to 5 
we can also see that with the increase of $\kappa$, which represents the
strength of confinement, the peaks of the heavy quark distribution functions 
become lower and their widths bigger. This is because the heavy quark
behaves more freely when confinement is weaker.

In Ref. \cite{tony} Close and Thomas calculated the $u$-quark distribution 
function in the nucleon in the MIT bag model, where numerical results
for $\alpha u(\alpha)$ at $\nu^2=5 \sim 10$GeV$^2$ were shown for various
values of the 
bag radius. It was shown that $\alpha u(\alpha)$ has a maximum value around
$0.4 \sim 0.5$ at $\alpha=0.4 \sim 0.5$. In general their argument predicts
the peak of the valence distribution (for ground state ($L=0$) baryons)
at $(m_H-m_D)/m_H$ (where $m_H$ is the baryon's mass), which is in the
range $0.87 \sim 0.88$ for $\Lambda_b$ and $0.67 \sim 0.72$ for $\Lambda_c$,
respectively. Clearly these expectations are in excellent agreement with
the peak positions at the hadronic scale, $\nu_0^2$. This statement is
also true for $\omega_Q$. We see clearly that with the increase of quark 
masses, the peak positions of the quark distribution functions move closer to 1
and the peak values become bigger.

In Tables 1,2,3,4 for different heavy baryons with different parameters in 
our model, we list the values of $\alpha_0$, the widths of these peaks,
and the total ``plus''
momentum fraction, $f$, carried by the heavy quark which is defined as 
\begin{equation}
f \equiv \int {\rm d}\alpha \alpha Q(\alpha). 
\label{3a0}
\end{equation}
The width is presented by two values of $\alpha$ at which $Q(\alpha)$ is 
half of its value at the peak, $Q(\alpha_0)$.
From these tables we can see the following points: (i) The values of
$\alpha_0$, $f$, and the widths depend not only on the light diquark mass $m_D$
but also on $\kappa$, and mostly the dependence on $\kappa$
is more sensitive for $c$-baryons.
(ii) The value of $\alpha_0$ decreases with the increase of $m_D$, which
is reasonable. (iii) The widths of heavy quark distribution
functions for $b$-baryons are much smaller than those for $c$-baryons. 
(iv) For $\Lambda_Q$, $1/m_Q$ corrections make $\alpha_0$ and $f$ bigger.
Hence the heavy quark carries more momentum with $1/m_Q$ corrections included.
(v) Numerically, in the whole range of parameters we are considering,
for $\Lambda_b$, $\alpha_0$ varies from 0.864 to 0.890 (0.870 to 0.890)
and $f$ varies from 0.800 to 0.857 (0.810 to 0.860) without (with) $1/m_Q$ 
corrections, while for $\Lambda_c$, $\alpha_0$ varies from 0.670 to 0.717 
(0.690 to 0.760) and $f$ varies from 0.522 to 0.651 (0.585 to 0.672);
for $\Sigma_b$, $\alpha_0$ varies from 0.823 to 0.846 
and $f$ varies from 0.809 to 0.834, while for $\Sigma_c$, $\alpha_0$ varies 
from 0.583 to 0.647 and $f$ varies from 0.493 to 0.589;
for $\Xi_b$, $\alpha_0$ varies from 0.803 to 0.820 
and $f$ varies from 0.780 to 0.809, while for $\Xi_c$, $\alpha_0$ varies 
from 0.503 to 0.580 and $f$ varies from 0.412 to 0.520;
for $\Omega_b$, $\alpha_0$ varies from 0.797 to 0.813 
and $f$ varies from 0.770 to 0.802, while for $\Omega_c$, $\alpha_0$ varies 
from 0.533 to 0.580 and $f$ varies from 0.449 to 0.537.

\begin{table}
\caption{Values  of  $\alpha_0$, widths, and $f$  
for $\Lambda_{b}$ with and without $1/m_Q$ corrections}
\begin{center}
\begin{tabular}{lcccc}
\hline
\hline
& &$m_D$=0.65GeV &  & \\ 
\hline
$\kappa$(GeV$^3$)&0.02 ($O(1)$)&0.02 ($O(1/m_Q)$)&0.10 ($O(1)$) 
&0.10 ($O(1/m_Q)$)\\
\hline
$\alpha_0$ &0.890 &0.890 &0.890 &0.890 \\
\hline
width &0.797 $\sim$ 0.923&0.833 $\sim$ 0.923
&0.773 $\sim$ 0.940&0.783 $\sim$ 0.943\\
\hline
$f$ &0.857&0.860&0.822&0.830\\
\hline
\hline
& &$m_D$=0.70GeV &  & \\ 
\hline
$\kappa$(GeV$^3$)&0.02 ($O(1)$)&0.02 ($O(1/m_Q)$)&0.10 ($O(1)$) 
&0.10 ($O(1/m_Q)$)\\
\hline
$\alpha_0$ &0.876 &0.878 &0.878 &0.884 \\
\hline
width &0.814 $\sim$ 0.918&0.816 $\sim$ 0.919
&0.762 $\sim$ 0.935&0.772 $\sim$ 0.937\\
\hline
$f$ &0.844&0.848&0.811&0.820\\
\hline
\hline
& &$m_D$=0.75GeV &  & \\ 
\hline
$\kappa$(GeV$^3$)&0.02 ($O(1)$)&0.02 ($O(1/m_Q)$)&0.10 ($O(1)$) 
&0.10 ($O(1/m_Q)$)\\
\hline
$\alpha_0$ &0.868 &0.870 &0.864 &0.870 \\
\hline
width &0.789 $\sim$ 0.912&0.802 $\sim$ 0.914
&0.750 $\sim$ 0.930&0.762 $\sim$ 0.932\\
\hline
$f$ &0.829&0.834&0.800&0.810\\
\hline
\hline
\end{tabular}
\end{center}
\end{table}

\begin{table}
\caption{Values  of  $\alpha_0$, widths, and $f$  
for $\Lambda_{c}$ with and without $1/m_Q$ corrections}
\begin{center}
\begin{tabular}{lcccc}
\hline
\hline
& &$m_D$=0.65GeV &  & \\ 
\hline
$\kappa$(GeV$^3$)&0.02 ($O(1)$)&0.02 ($O(1/m_Q)$)&0.10 ($O(1)$) 
&0.10 ($O(1/m_Q)$)\\
\hline
$\alpha_0$ &0.717 &0.730 &0.710 &0.760 \\
\hline
width &0.580 $\sim$ 0.807&0.600 $\sim$ 0.817
&0.443 $\sim$ 0.857&0.510 $\sim$ 0.870\\
\hline
$f$ &0.651&0.672&0.569&0.625\\
\hline
\hline
& &$m_D$=0.70GeV &  & \\ 
\hline
$\kappa$(GeV$^3$)&0.02 ($O(1)$)&0.02 ($O(1/m_Q)$)&0.10 ($O(1)$) 
&0.10 ($O(1/m_Q)$)\\
\hline
$\alpha_0$ &0.697 &0.717 &0.697 &0.730 \\
\hline
width &0.537 $\sim$ 0.797&0.563 $\sim$ 0.807
&0.417 $\sim$ 0.840&0.487 $\sim$ 0.856\\
\hline
$f$ &0.617&0.645&0.546&0.605\\
\hline
\hline
& &$m_D$=0.75GeV &  & \\ 
\hline
$\kappa$(GeV$^3$)&0.02 ($O(1)$)&0.02 ($O(1/m_Q)$)&0.10 ($O(1)$) 
&0.10 ($O(1/m_Q)$)\\
\hline
$\alpha_0$ &0.670 &0.690 &0.670 &0.713 \\
\hline
width &0.500 $\sim$ 0.783&0.530 $\sim$ 0.797
&0.387 $\sim$ 0.823&0.550 $\sim$ 0.810\\
\hline
$f$ &0.584&0.618&0.522&0.585\\
\hline
\hline
\end{tabular}
\end{center}
\end{table}

\begin{table}
\caption{Values  of  $\alpha_0$, widths, and $f$  
for $\Sigma_{b}$ and $\Sigma_{c}$ in the heavy quark limit}
\begin{center}
\begin{tabular}{lcccc}
\hline
\hline
& &$m_D$=0.90GeV &  & \\ 
\hline
$\kappa$(GeV$^3$)&0.02 ($\Sigma_{b}$)&0.10 ($\Sigma_{b}$)&0.02 ($\Sigma_{c}$) 
&0.10 ($\Sigma_{c}$)\\
\hline
$\alpha_0$ &0.846 &0.843 &0.630 &0.647 \\
\hline
width &0.800 $\sim$ 0.883&0.750 $\sim$ 0.907
&0.517 $\sim$ 0.720&0.397 $\sim$ 0.773\\
\hline
$f$ &0.834&0.834&0.589&0.528\\
\hline
\hline
& &$m_D$=0.95GeV &  & \\ 
\hline
$\kappa$(GeV$^3$)&0.02 ($\Sigma_{b}$)&0.10 ($\Sigma_{b}$)&0.02 ($\Sigma_{c}$) 
&0.10 ($\Sigma_{c}$)\\
\hline
$\alpha_0$ &0.840 &0.843 &0.610 &0.600 \\
\hline
width &0.780 $\sim$ 0.880&0.733 $\sim$ 0.903
&0.470 $\sim$ 0.720&0.463 $\sim$ 0.723\\
\hline
$f$ &0.820&0.813&0.550&0.493\\
\hline
\hline
& &$m_D$=1.00GeV &  & \\ 
\hline
$\kappa$(GeV$^3$)&0.02 ($\Sigma_{b}$)&0.10 ($\Sigma_{b}$)&0.02 ($\Sigma_{c}$) 
&0.10 ($\Sigma_{c}$)\\
\hline
$\alpha_0$ &0.830 &0.823 &0.583 &0.597 \\
\hline
width &0.763 $\sim$ 0.923&0.740 $\sim$ 0.889
&0.430 $\sim$ 0.707&0.383 $\sim$ 0.730\\
\hline
$f$ &0.809&0.814&0.527&0.523\\
\hline
\hline
\end{tabular}
\end{center}
\end{table}

\begin{table}
\caption{Values  of  $\alpha_0$, widths, and $f$  
for $\Xi_{b}$ and $\Xi_{c}$ in the heavy quark limit}
\begin{center}
\begin{tabular}{lcccc}
\hline
\hline
& &$m_D$=1.10GeV &  & \\ 
\hline
$\kappa$(GeV$^3$)&0.02 ($\Xi_{b}$)&0.10 ($\Xi_{b}$)&0.02 ($\Xi_{c}$) 
&0.10 ($\Xi_{c}$)\\
\hline
$\alpha_0$ &0.820 &0.817 &0.553 &0.580 \\
\hline
width &0.773 $\sim$ 0.857&0.733 $\sim$ 0.877
&0.440 $\sim$ 0.647&0.343 $\sim$ 0.700\\
\hline
$f$ &0.809&0.796&0.520&0.471\\
\hline
\hline
& &$m_D$=1.15GeV &  & \\ 
\hline
$\kappa$(GeV$^3$)&0.02 ($\Xi_{b}$)&0.10 ($\Xi_{b}$)&0.02 ($\Xi_{c}$) 
&0.10 ($\Xi_{c}$)\\
\hline
$\alpha_0$ &0.810 &0.810 &0.533 &0.510 \\
\hline
width &0.753 $\sim$ 0.850&0.717 $\sim$ 0.875
&0.393 $\sim$ 0.643&0.300 $\sim$ 0.690\\
\hline
$f$ &0.795&0.791&0.485&0.440\\
\hline
\hline
& &$m_D$=1.20GeV &  & \\ 
\hline
$\kappa$(GeV$^3$)&0.02 ($\Xi_{b}$)&0.10 ($\Xi_{b}$)&0.02 ($\Xi_{c}$) 
&0.10 ($\Xi_{c}$)\\
\hline
$\alpha_0$ &0.803 &0.803 &0.513 &0.503 \\
\hline
width &0.737 $\sim$ 0.853&0.700 $\sim$ 0.870
&0.347 $\sim$ 0.640&0.257 $\sim$ 0.677\\
\hline
$f$ &0.782&0.780&0.445&0.412\\
\hline
\hline
\end{tabular}
\end{center}
\end{table}

\begin{table}
\caption{Values  of  $\alpha_0$, widths, and $f$  
for $\Omega_{b}$ and $\Omega_{c}$ in the heavy quark limit}
\begin{center}
\begin{tabular}{lcccc}
\hline
\hline
& &$m_D$=1.15GeV &  & \\ 
\hline
$\kappa$(GeV$^3$)&0.02 ($\Omega_{b}$)&0.10 ($\Omega_{b}$)&0.03 ($\Omega_{c}$) 
&0.10 ($\Omega_{c}$)\\
\hline
$\alpha_0$ &0.813 &0.813 &0.573 &0.580 \\
\hline
width &0.760 $\sim$ 0.853&0.727 $\sim$ 0.873
&0.453 $\sim$ 0.667&0.373 $\sim$ 0.710\\
\hline
$f$ &0.802&0.772&0.537&0.460\\
\hline
\hline
& &$m_D$=1.20GeV &  & \\ 
\hline
$\kappa$(GeV$^3$)&0.02 ($\Omega_{b}$)&0.10 ($\Omega_{b}$)&0.02 ($\Omega_{c}$) 
&0.10 ($\Omega_{c}$)\\
\hline
$\alpha_0$ &0.803 &0.810 &0.553 &0.557 \\
\hline
width &0.750 $\sim$ 0.894&0.710 $\sim$ 0.867
&0.430 $\sim$ 0.650&0.343 $\sim$ 0.697\\
\hline
$f$ &0.786&0.774&0.510&0.464\\
\hline
\hline
& &$m_D$=1.25GeV &  & \\ 
\hline
$\kappa$(GeV$^3$)&0.02 ($\Omega_{b}$)&0.10 ($\Omega_{b}$)&0.02 ($\Omega_{c}$) 
&0.10 ($\Omega_{c}$)\\
\hline
$\alpha_0$ &0.797 &0.797 &0.533 &0.540 \\
\hline
width &0.733 $\sim$ 0.847&0.697 $\sim$ 0.863
&0.393 $\sim$ 0.650&0.313 $\sim$ 0.687\\
\hline
$f$ &0.773&0.770&0.474&0.449\\
\hline
\hline
\end{tabular}
\end{center}
\end{table}

As mentioned in Section II, the heavy quark distribution functions should
be normalized according to Eq. (\ref{2h1}).
For some fixed set of parameters in our model, the B-S wave functions can be
completely determined by solving the B-S equations numerically and fixing the
normalization of the corresponding Isgur-Wise functions at the zero-recoil
point. Hence by checking whether the normalization condition Eq. (\ref{2h1})
is satisfied or not we can test whether our model works well. We have performed
the integration of $Q(\alpha)$ over $\alpha$ numerically. It is found that 
for $\Lambda_b$ and $\Lambda_c$, in the whole range of the parameters, 
$\int {\rm d}\alpha Q(\alpha)$ is very close to 1 (for $\Lambda_b$ 
and $\Lambda_c$ the deviation from 1 is of order $10^{-4}$ and
$10^{-3} \sim 10^{-2}$, respectively). Furthermore, we find that $1/m_Q$
corrections make $\int {\rm d}\alpha Q(\alpha)$ even closer to 1. Therefore,
our model for $1/m_Q$ corrections works in the right direction. For
$\omega_b$ and $\omega_c$, although $\int {\rm d}\alpha Q(\alpha)$  is not 
as close to 1 as in the case of $\Lambda_Q$, the deviation from 1 still
happens at order $10^{-2}$ at most, which is expected to be improved if
$1/m_Q$ corrections are also included. Thus the momentum sum rule is
satisfied in our model.

Now we compare the present results with another phenomenological model.
In Ref. \cite{guo}, also based on the quark-diquark picture, $\Lambda_Q$ 
is regarded as composed of a heavy quark and a scalar light diquark.
The heavy baryon wave function $\Psi_{\Lambda_{Q}}(\alpha,{\bf k}_{\perp})$
is proposed as a generalization of the Bauer-Stech-Wirbel \cite{bsw} meson
wave function to the quark-diquark case,
\begin{equation}
\Psi_{\Lambda_{Q}}(x_{1},{\bf k}_{\perp}) = N_{\Lambda_{Q}}\alpha (1-\alpha)^3
{\rm exp}\{-b^{2}[{\bf k}_{\perp}^2+
m_{\Lambda_{Q}}^{2}(\alpha-\bar{\alpha})^{2}]\},
\label{3a}
\end{equation}
where $N_{\Lambda_{Q}}$ is the normalization constant,
$\bar{\alpha}=m_Q/m_{\Lambda_{Q}}$, 
${\bf k}_{\perp}$ is the transverse momentum
and the parameter $b$ is related to the root of the average square of 
${\bf k}_{\perp}$, 
$\sqrt{\langle {\bf k}^{2}_{\perp} \rangle}$. The normalization of the 
wave function is
\begin{equation}
\int{\rm d}\alpha {\rm d}^{2}{\bf k}_{\perp}|\Psi_{\Lambda_{Q}}(\alpha,
{\bf k}_{\perp})|^{2} = 1. 
\label{3b}
\end{equation}
 
The heavy quark distribution function in this model is defined in the
following
\begin{equation}
Q(\alpha)=\int {\rm d}^{2}{\bf k}_{\perp}|\Psi_{\Lambda_{Q}}(\alpha,
{\bf k}_{\perp})|^{2},
\label{3c}
\end{equation}
which leads to
\begin{equation}
Q(\alpha)=\frac{2\pi^2 N_{\Lambda_{Q}}^2}{b^2} \alpha^2 (1-\alpha)^6
{\rm exp}[-2b^{2}m_{\Lambda_{Q}}^{2}(\alpha-\bar{\alpha})^{2}].
\label{3c1}
\end{equation}

Here $\langle {\bf k}^{2}_{\perp} \rangle$ is defined as
\begin{equation}
\langle {\bf k}^{2}_{\perp} \rangle=\int {\rm d}\alpha {\rm d}^{2}
{\bf k}_{\perp}{\bf k}_{\perp}^{2} 
|\Psi_{\Lambda_{Q}}(\alpha, {\bf k}_{\perp})|^{2},
\label{3d}
\end{equation}
from which, using Eq. (\ref{3b}), we have 
\begin{equation}
2b^2 \langle {\bf k}^{2}_{\perp} \rangle=1.
\label{3e}
\end{equation}
 
We choose $\sqrt{\langle {\bf k}^{2}_{\perp} \rangle}$ to vary between
400MeV and 600MeV. For these two parameters we plot $Q(\alpha)$ at the hadronic
scale in Fig.6(a) and we plot 
$\alpha Q(\alpha)$ at $\nu^2=10$GeV$^2$ in Fig.6(b). 
We can see from these plots that the heavy quark distribution 
functions in this model have a similar shape to those found in the present
B-S model. The peak position for $Q(\alpha)$, 
widths of the distributions, and the total ``plus'' 
momentum fraction carried by the
heavy quark $Q$, are listed in Table 6. Notice that
these values are independent of the value of $m_D$ in this model.
Finally, it can be seen from Table 6 that both $\alpha_0$ and $f$ in 
this model are smaller than those in the B-S model.

\begin{table}
\caption{Values  of  $\alpha_0$, widths, and $f$  
for $\Lambda_{b}$ and $\Lambda_{c}$ in the model of Ref. \cite{guo}}
\begin{center}
\begin{tabular}{lcccc}
\hline
\hline
$\sqrt{\langle {\bf k}^{2}_{\perp} \rangle}$(GeV)&0.4 ($\Lambda_{b}$)&0.6 
($\Lambda_{b}$)&0.4 ($\Lambda_{c}$) 
&0.6 ($\Lambda_{c}$)\\
\hline
$\alpha_0$ &0.816 &0.764 &0.545 &0.460 \\
\hline
width &0.766 $\sim$ 0.864&0.693 $\sim$ 0.830
&0.426 $\sim$ 0.660&0.306 $\sim$ 0.609\\
\hline
$f$ &0.813&0.758&0.540&0.456\\
\hline
\hline
\end{tabular}
\end{center}
\end{table}

\vspace{0.2in}
{\large\bf IV. Summary and discussion}
\vspace{0.2in}

Based on our heavy quark and light diquark model for heavy baryons $\Lambda_Q$
and $\omega_Q$ in the B-S equation approach, we have calculated the heavy quark
distribution functions in these baryons. This was done
in the heavy quark limit for both $\Lambda_Q$ and $\omega_Q$. Furthermore,
$1/m_Q$ corrections were also included in the case of $\Lambda_Q$.  
Our numerical results show that for different heavy baryons with the same 
heavy quark flavor the shapes of the heavy quark distribution functions 
are similar. At the hadronic scale, $\Lambda_{\rm QCD}$, they have an obvious
peak at some ``plus'' momentum fraction carried by the heavy quark, $\alpha_0$,
which is much closer to 1 for $b$-baryons than $c$-baryons. In addition, the
widths of these distributions are much smaller for $b$-baryons than 
$c$-baryons. Furthermore, we have found that the peaks of the heavy quark 
distribution functions become lower and their widths bigger when 
the confinement between the heavy quark and light diquark inside a
heavy baryon is stronger.  
The QCD evolution of these distribution functions were also 
discussed by applying DGLAP equations to the next-to-leading order and
the results show that the distinction between $b$-quark and $c$-quark
distribution functions is still obvious at high $\nu^2$. The
evolution makes the amplitudes of the peaks much smaller, especially for
the distribution functions in $b$-baryons. We also calculated the total
``plus'' momentum 
fraction carried by the heavy quark, which is much bigger for 
$b$-baryons than $c$-baryons. The dependence of all these numerical results
on the parameters in our model was discussed in detail. We also checked
the normalization condition of these distribution functions and found
that it is satisfied. In addition, we
compared our results with an existing phenomenological model for
$\Lambda_Q$ proposed in Ref. \cite{guo}. It was found that the shapes 
of the heavy quark distribution functions in these two models are quite
similar, although the peak positions in the B-S model are closer to
1 than those in the model of Ref. \cite{guo}.
We also compared our results for the heavy quark distribution functions with 
the general argument of Close and Thomas based on energy and momentum
conservation. It was found that the peak positions of the quark 
distribution functions move closer to 1 and the peak values become bigger
when the quark masses increase. 

Although we cannot determine the exact value of $\nu_0$, the hadronic scale
at which the
heavy quark distribution functions are calculated, our results do reflect
nonperturbative information in the heavy baryons $\Lambda_Q$ and $\omega_Q$. 
We have treated $\nu_0$ as a parameter to be determined by experiment.
Even though the heavy quark distribution functions cannot be measured 
directly from deep inelastic scattering processes,
we can still expect to obtain some information about these functions 
from the decays of heavy baryons and fragmentation of heavy quarks to
heavy baryons. In fact, the latter is directly related to experimental
measurements of fragmentation processes. However, the theoretical extension
to address fragmentation processes is more complicated. This is currently
under investigation. 

\vspace{2cm}

\noindent {\bf Acknowledgment}:

One of us (Guo) gratefully acknowledges F.-G. Cao for helpful discussions. 
He also  thanks O. Leitner and F. Bissey for assistance in preparing the 
figures. This work was supported by the Australian Research Council.

\newpage

\vspace{0.2in}

\noindent{\large \bf Figure Captions} \\
\vspace{0.4cm}

\noindent Fig.1 (a) Heavy quark distribution functions at the hadronic 
scale $\nu_0^2$ with the B-S equation solved for $\Lambda_Q$ in the limit
$m_Q \rightarrow \infty$; (b) $\alpha Q(\alpha)$ for $\Lambda_Q$ at 
$\nu^2=10$GeV$^2$ in the limit $m_Q \rightarrow \infty$. The lines
on the right (left) are for $\Lambda_b$ ($\Lambda_c$). The solid (dot)
lines correspond to $m_D=0.70$GeV and $\kappa=0.02$GeV$^3$ 
($\kappa=0.10$GeV$^3$). The dashed (dot dashed) lines correspond to
$\kappa=0.06$GeV$^3$ and $m_D=0.65$GeV ($m_D=0.75$GeV). 

\vspace{0.2cm}

\noindent Fig.2 (a) Heavy quark distribution functions at the hadronic scale 
$\nu_0^2$ with the B-S
equation solved for $\Lambda_Q$ to 
order $1/m_Q$; (b) $\alpha Q(\alpha)$ for $\Lambda_Q$ at $\nu^2=10$GeV$^2$
to order $1/m_Q$. The lines
on the right (left) are for $\Lambda_b$ ($\Lambda_c$). The solid (dot)
lines correspond to $m_D=0.70$GeV and $\kappa=0.02$GeV$^3$ 
($\kappa=0.10$GeV$^3$). The dashed (dot dashed) lines correspond to
$\kappa=0.06$GeV$^3$ and $m_D=0.65$GeV ($m_D=0.75$GeV).

\vspace{0.2cm}

\noindent Fig.3 (a) Heavy quark distribution functions 
at the hadronic scale $\nu_0^2$ with the B-S
equation solved for $\Sigma_Q$ in the limit
$m_Q \rightarrow \infty$; (b) $\alpha Q(\alpha)$ for $\Sigma_Q$ 
at $\nu^2=10$GeV$^2$ in the limit $m_Q \rightarrow \infty$. The lines
on the right (left) are for $\Sigma_b$ ($\Sigma_c$). The solid (dot)
lines correspond to $m_D=0.95$GeV and $\kappa=0.02$GeV$^3$ 
($\kappa=0.10$GeV$^3$). The dashed (dot dashed) lines correspond to
$\kappa=0.06$GeV$^3$ and $m_D=0.90$GeV ($m_D=1.00$GeV).

\vspace{0.2cm}

\noindent Fig.4 (a) Heavy quark distribution functions 
at the hadronic scale $\nu_0^2$ with the B-S
equation solved for $\Xi_Q$ 
in the limit $m_Q \rightarrow \infty$; (b) $\alpha Q(\alpha)$ for $\Xi_Q$ at 
$\nu^2=10$GeV$^2$
in the limit $m_Q \rightarrow \infty$. The lines
on the right (left) are for $\Xi_b$ ($\Xi_c$). The solid (dot)
lines correspond to $m_D=1.15$GeV and $\kappa=0.02$GeV$^3$ 
($\kappa=0.10$GeV$^3$). The dashed (dot dashed) lines correspond to
$\kappa=0.06$GeV$^3$ and $m_D=1.10$GeV ($m_D=1.20$GeV).

\vspace{0.2cm}

\noindent Fig.5 (a) Heavy quark distribution functions 
at the hadronic scale $\nu_0^2$ with the B-S
equation solved for $\Omega_Q$ 
in the limit 
$m_Q \rightarrow \infty$; (b) $\alpha Q(\alpha)$ for $\Omega_Q$ at 
$\nu^2=10$GeV$^2$ in the limit $m_Q \rightarrow \infty$. The lines
on the right (left) are for $\Omega_b$ ($\Omega_c$). The solid (dot)
lines correspond to $m_D=1.20$GeV and $\kappa=0.02$GeV$^3$ 
($\kappa=0.10$GeV$^3$). The dashed (dot dashed) lines correspond to
$\kappa=0.06$GeV$^3$ and $m_D=1.15$GeV ($m_D=1.25$GeV).

\vspace{0.2cm}

\noindent Fig.6 (a) Heavy quark distribution functions for $\Lambda_Q$ 
in the model of Ref. \cite{guo}; (b)$\alpha Q(\alpha)$ for $\Lambda_Q$ at 
$\nu^2=10$GeV$^2$ in the model of Ref. \cite{guo}.
The two lines on the right (left) are 
for $\Lambda_b$ ($\Lambda_c$). The solid (dashed)
lines correspond to $\sqrt{\langle k_{\perp} \rangle^2}=0.4$GeV 
($\sqrt{\langle k_{\perp} \rangle^2}=0.6$GeV).

\begin{figure}[p]
\begin{center}
{\epsfsize=14.7in\epsfbox{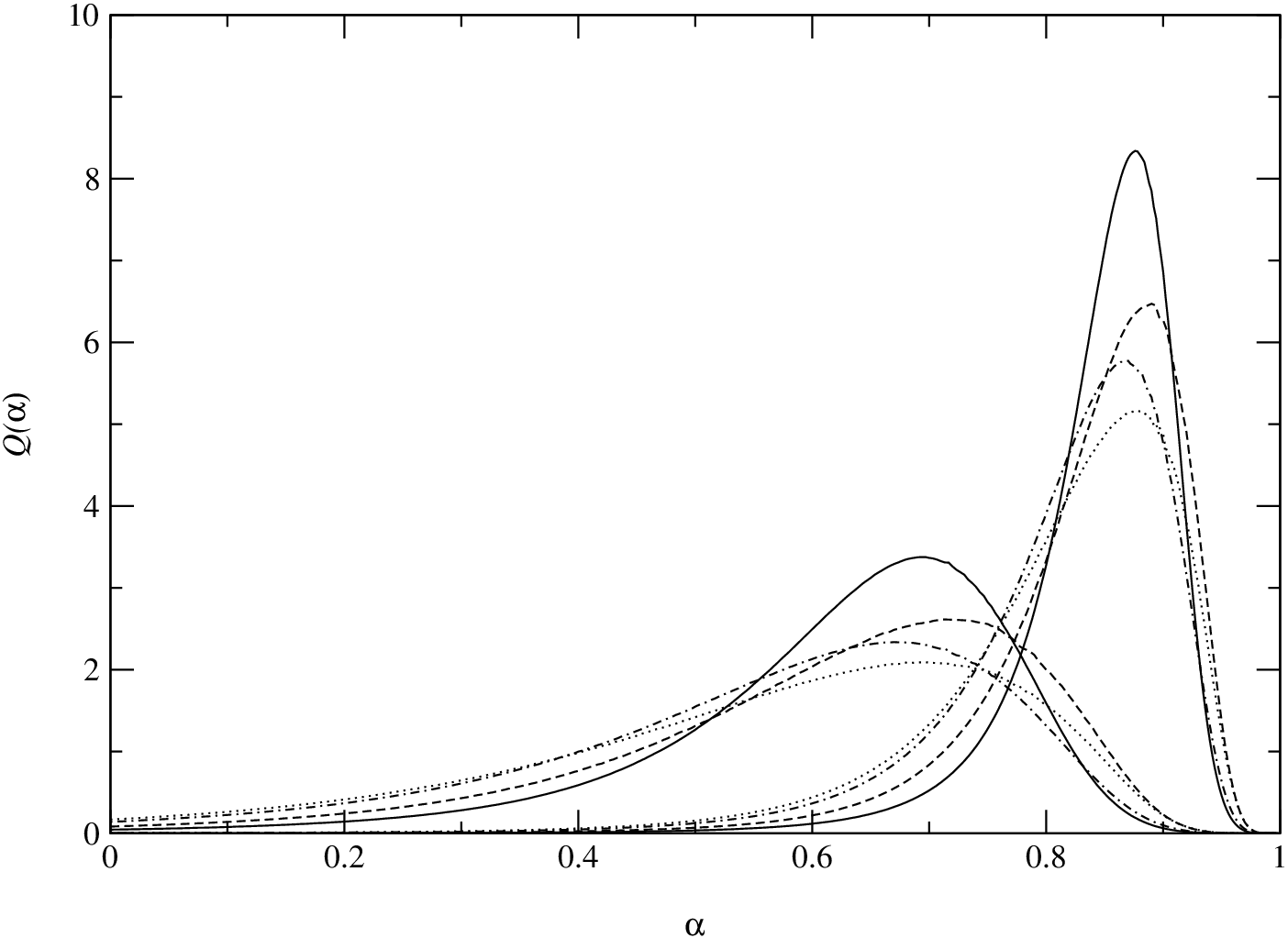}}
\end{center}
\vspace{0.3cm}
\centerline{Fig.1(a)}
\end{figure}

\begin{figure}[p]
\begin{center}
{\epsfsize=14.7in\epsfbox{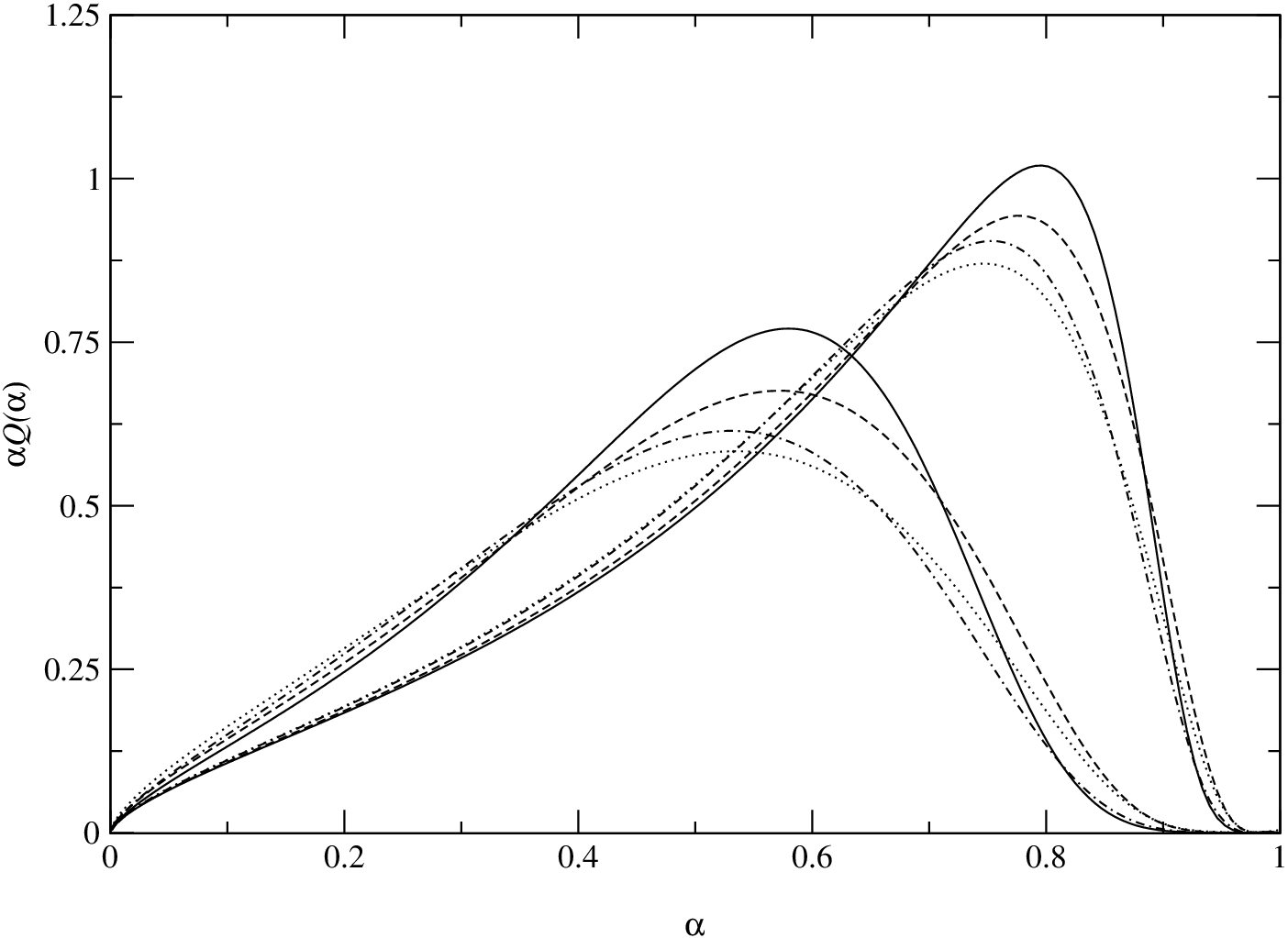}}
\end{center}
\vspace{0.3cm}
\centerline{Fig.1(b)}
\end{figure}

\begin{figure}[p]
\begin{center}
{\epsfsize=14.7in\epsfbox{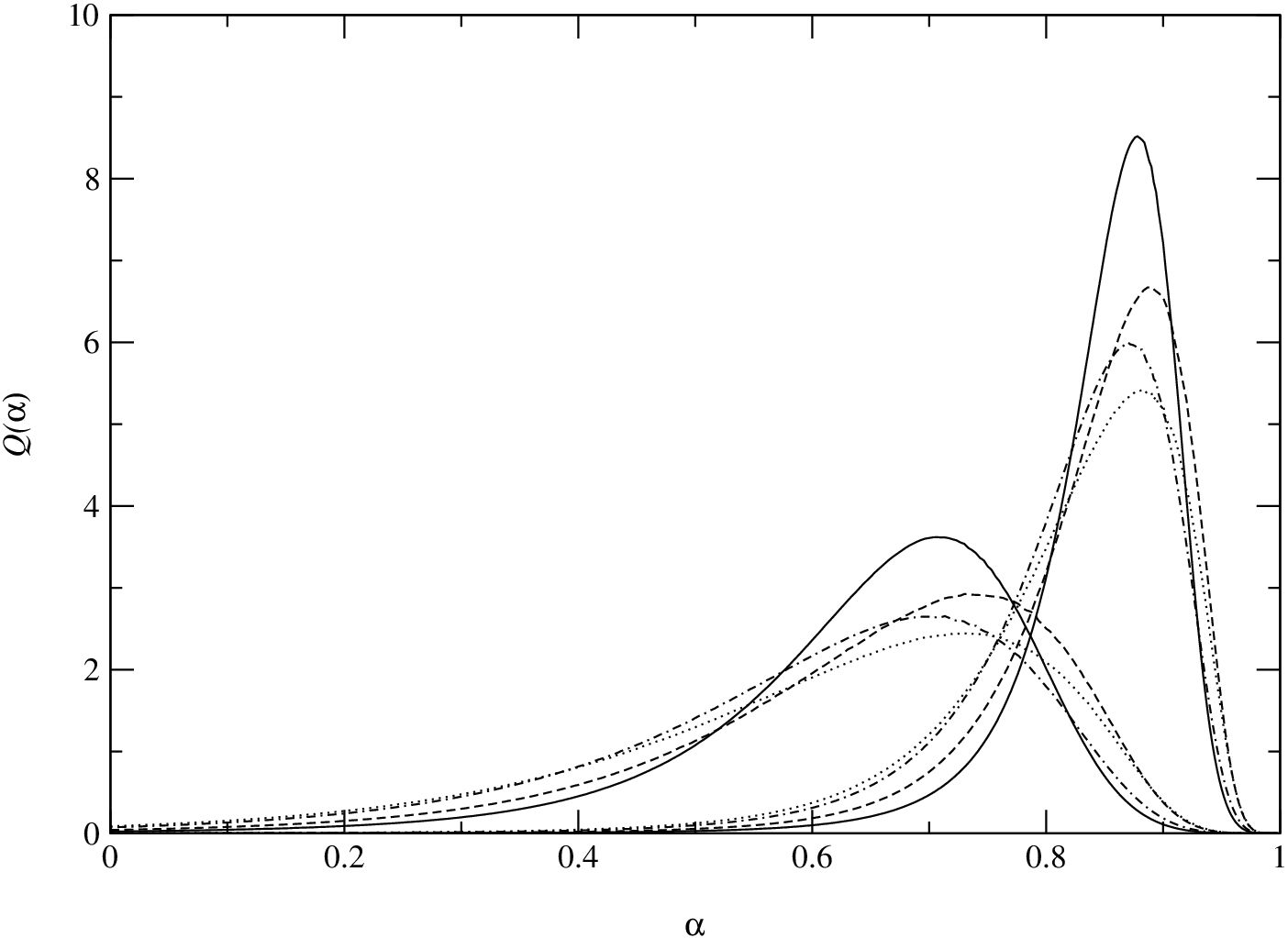}}
\end{center}
\vspace{0.3cm}
\centerline{Fig.2(a)}
\end{figure}

\begin{figure}[p]
\begin{center}
{\epsfsize=14.7in\epsfbox{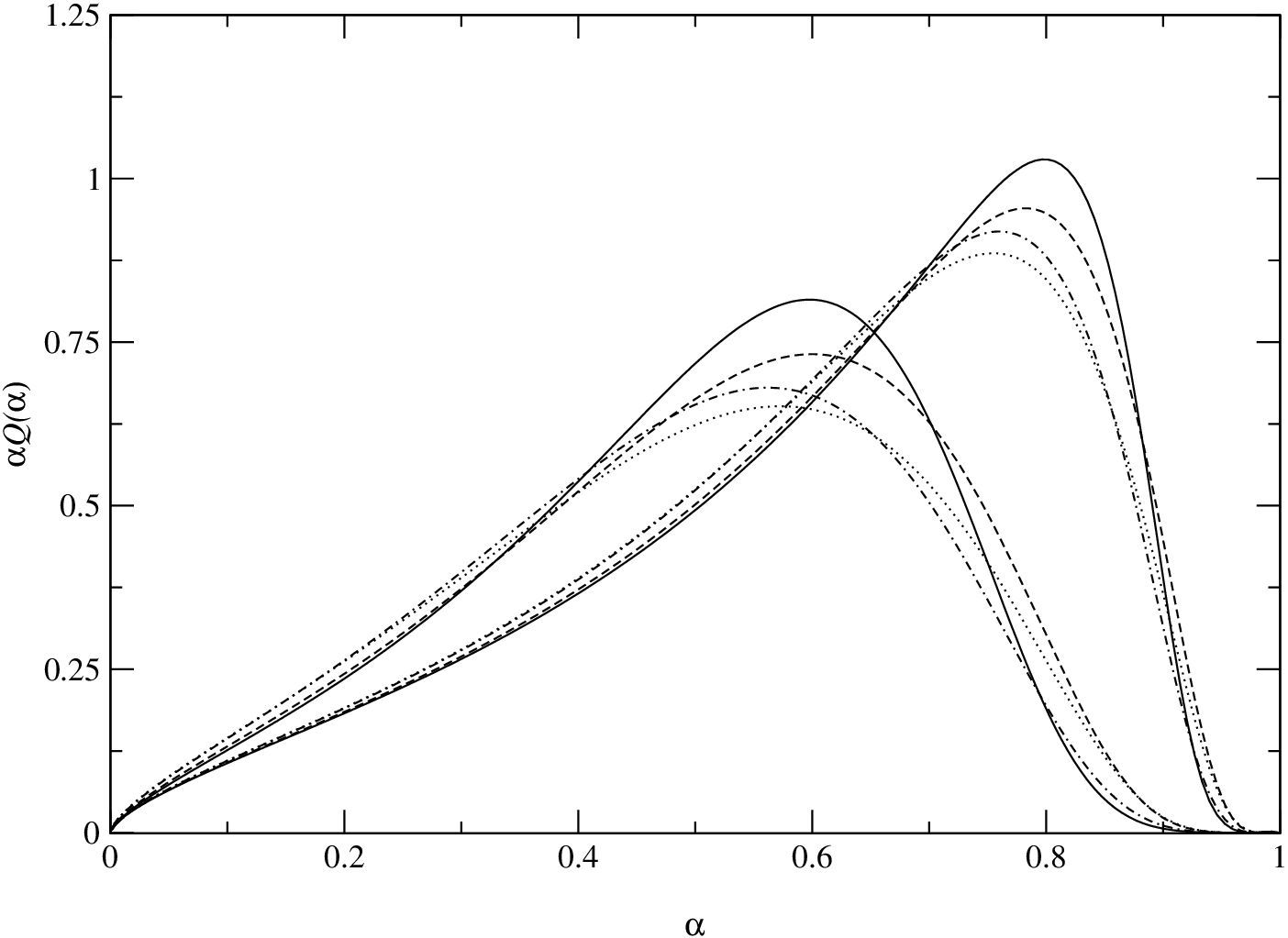}}
\end{center}
\vspace{0.3cm}
\centerline{Fig.2(b)}
\end{figure}

\begin{figure}[p]
\begin{center}
{\epsfsize=14.7in\epsfbox{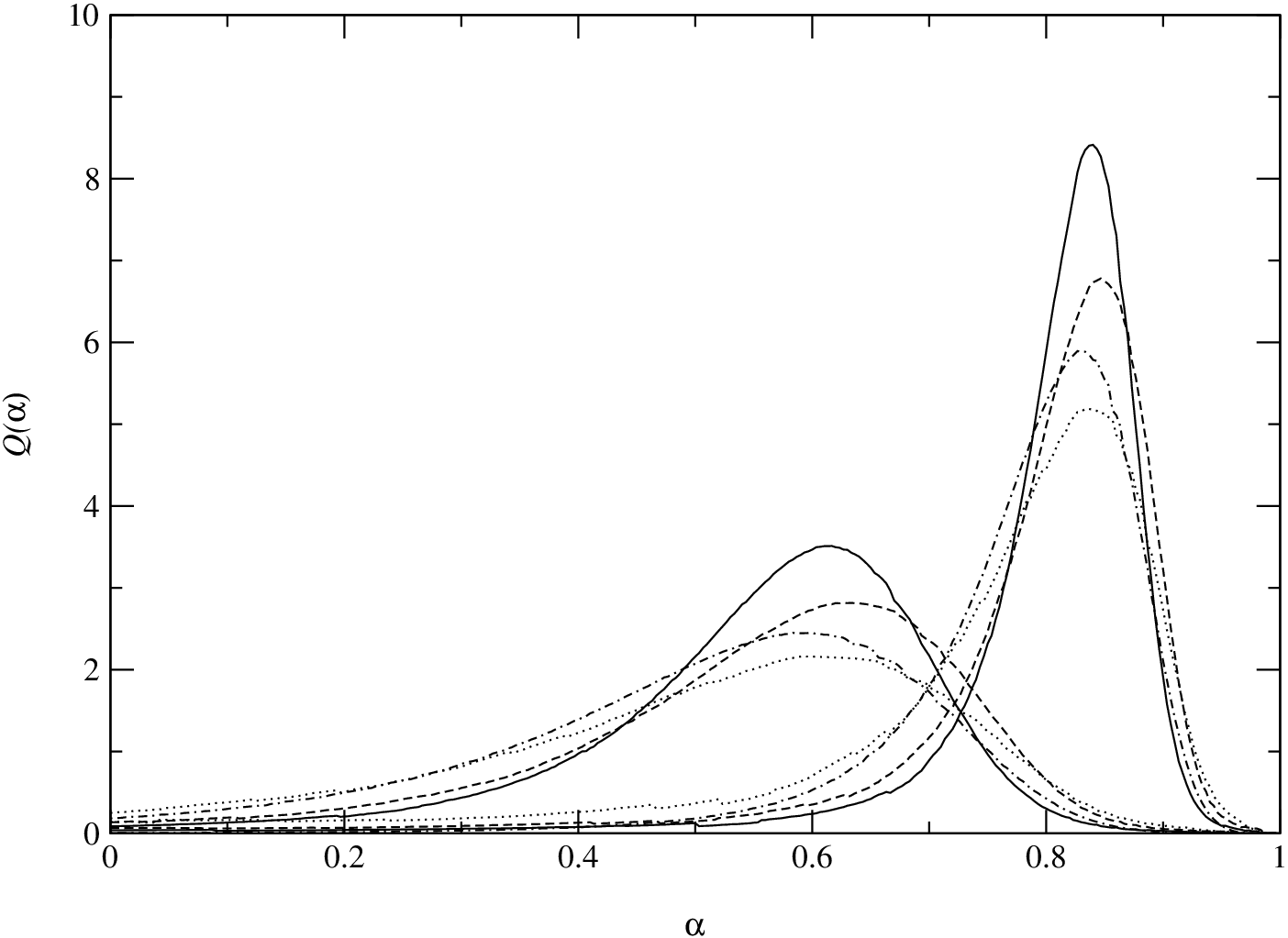}}
\end{center}
\vspace{0.3cm}
\centerline{Fig.3(a)}
\end{figure}

\begin{figure}[p]
\begin{center}
{\epsfsize=14.7in\epsfbox{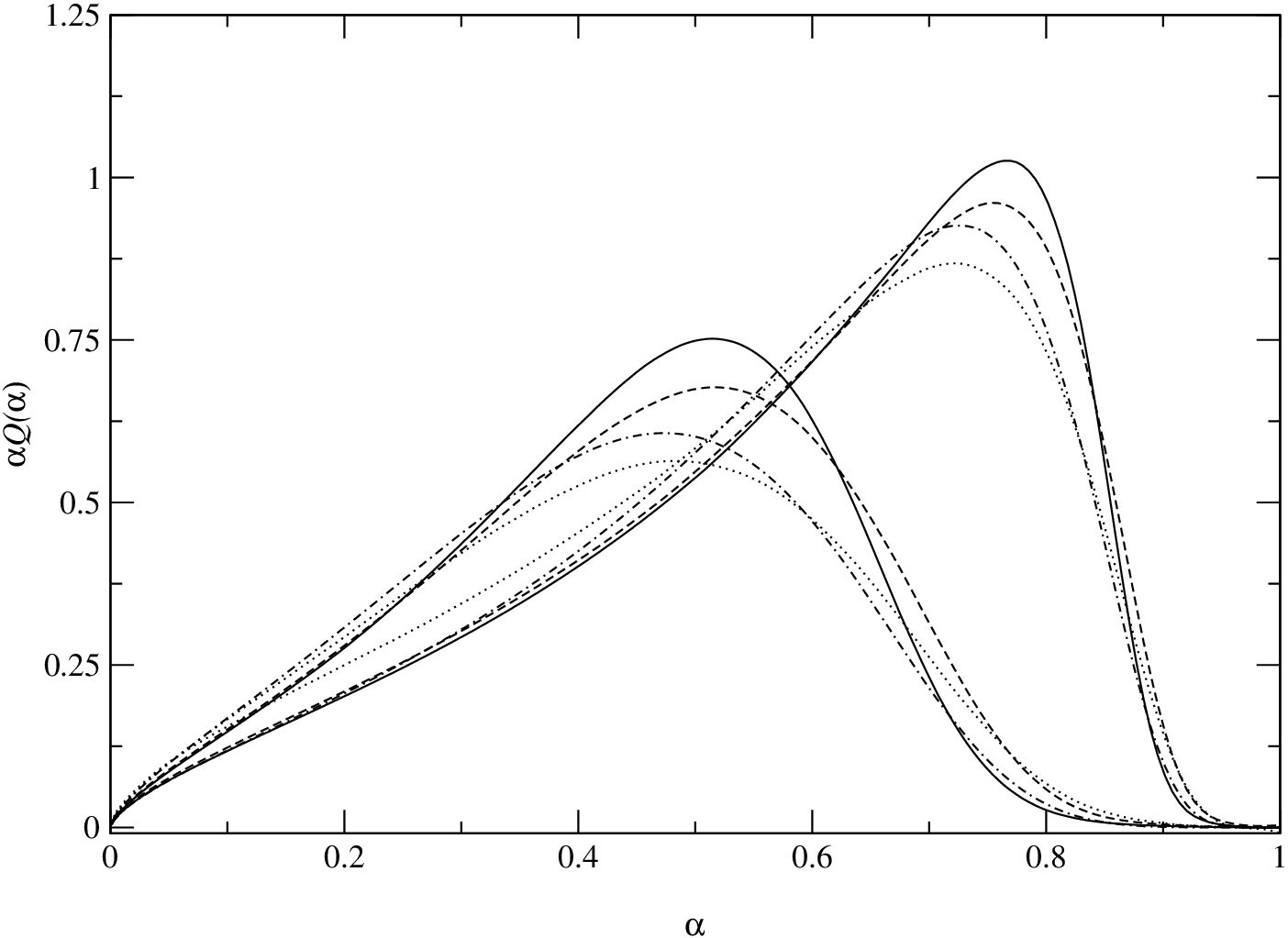}}
\end{center}
\vspace{0.3cm}
\centerline{Fig.3(b)}
\end{figure}

\begin{figure}[p]
\begin{center}
{\epsfsize=14.7in\epsfbox{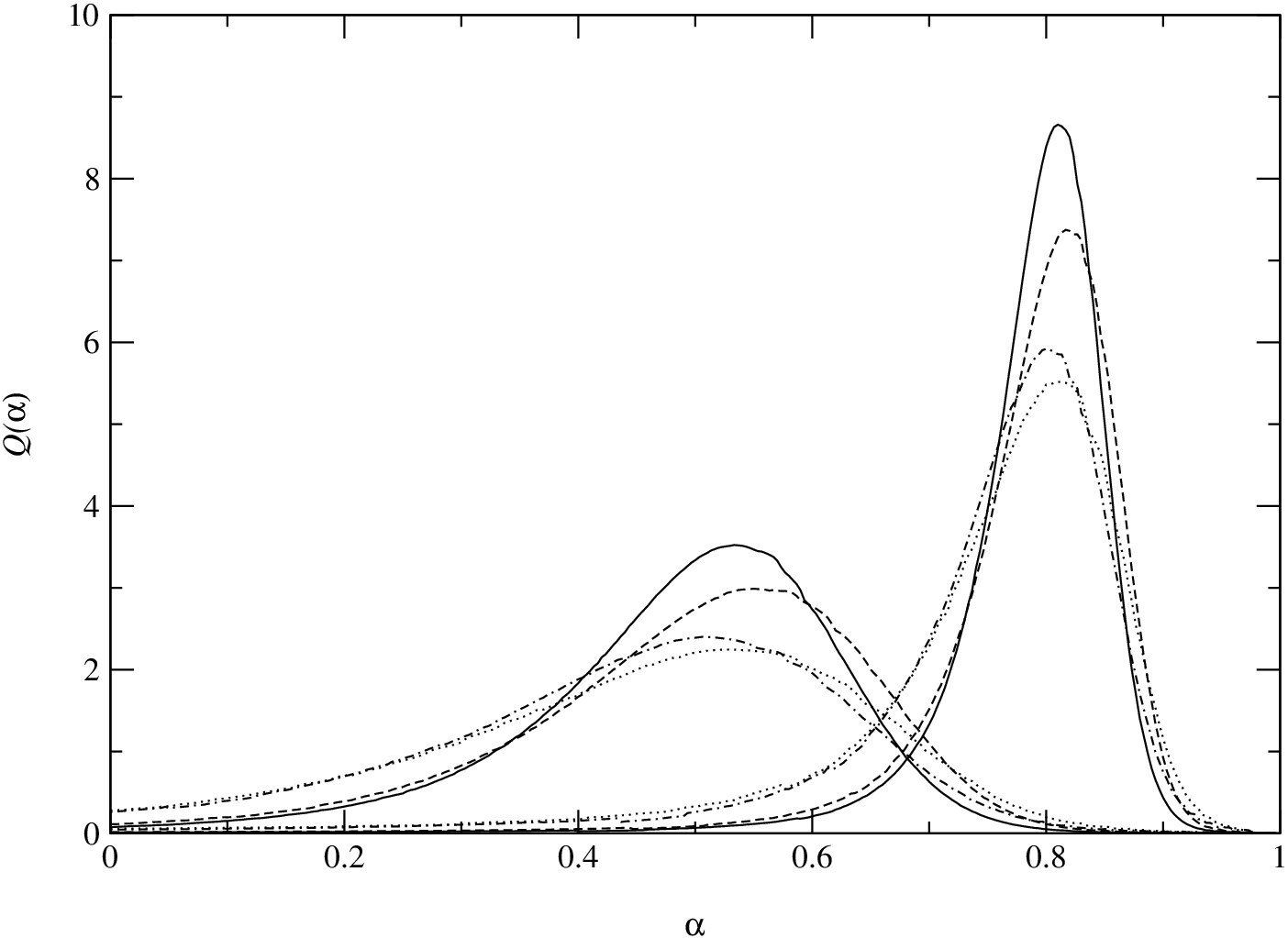}}
\end{center}
\vspace{0.3cm}
\centerline{Fig.4(a)}
\end{figure}

\begin{figure}[p]
\begin{center}
{\epsfsize=14.7in\epsfbox{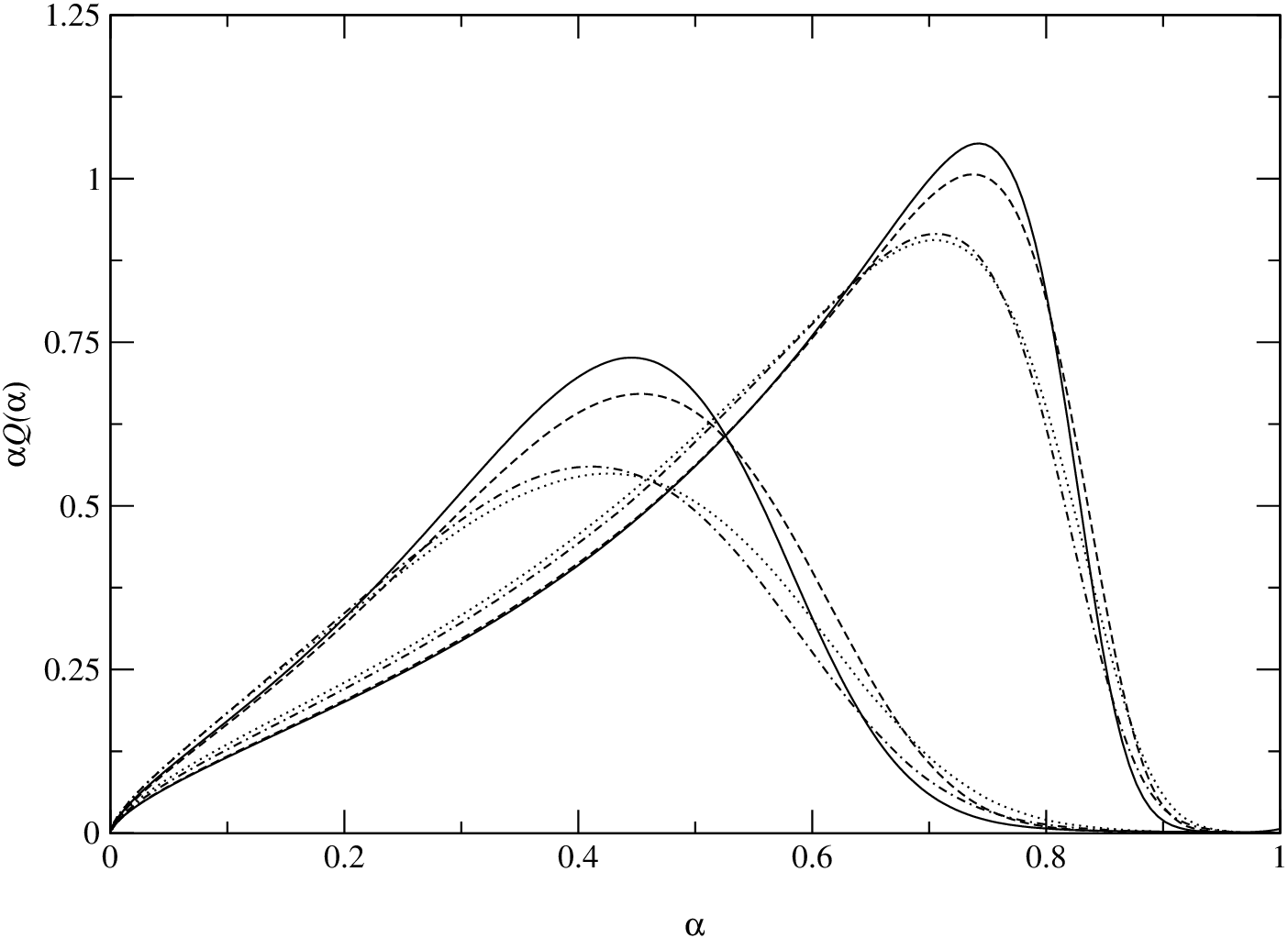}}
\end{center}
\vspace{0.3cm}
\centerline{Fig.4(b)}
\end{figure}

\begin{figure}[p]
\begin{center}
{\epsfsize=14.7in\epsfbox{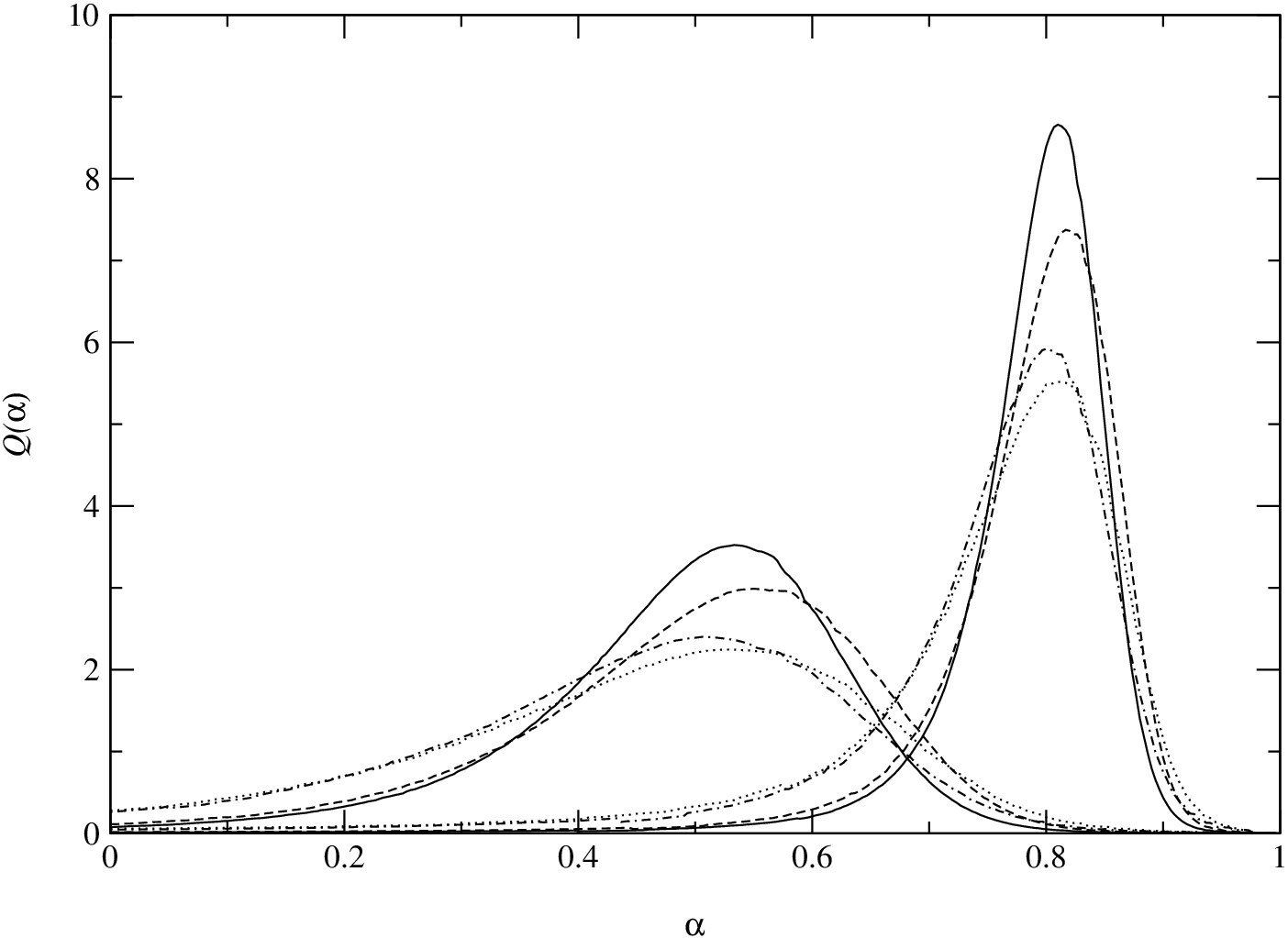}}
\end{center}
\vspace{0.3cm}
\centerline{Fig.5(a)}
\end{figure}

\begin{figure}[p]
\begin{center}
{\epsfsize=14.7in\epsfbox{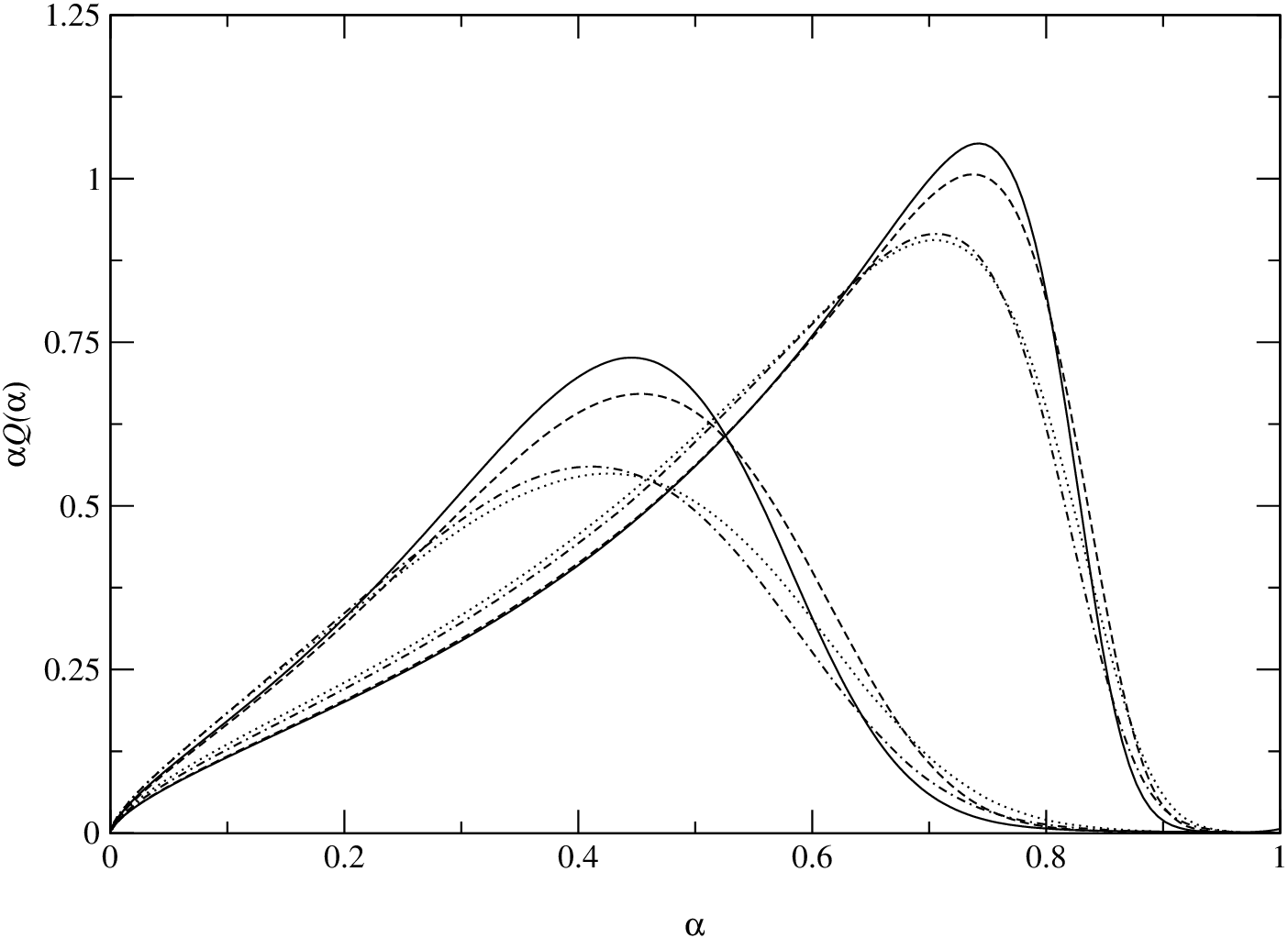}}
\end{center}
\vspace{0.3cm}
\centerline{Fig.5(b)}
\end{figure}

\begin{figure}[p]
\begin{center}
{\epsfsize=14.7in\epsfbox{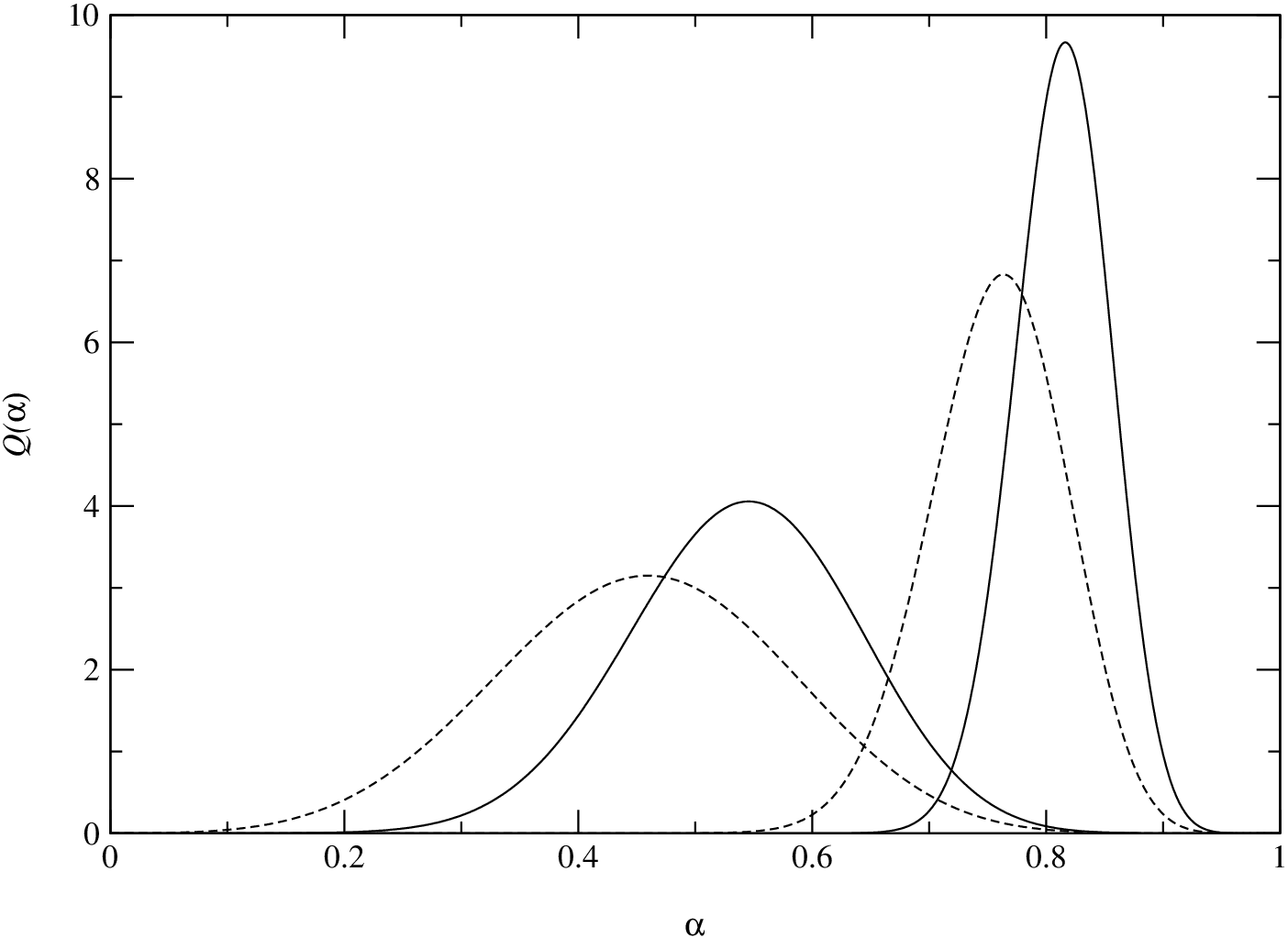}}
\end{center}
\vspace{0.3cm}
\centerline{Fig.6(a)}
\end{figure}

\begin{figure}[p]
\begin{center}
{\epsfsize=14.7in\epsfbox{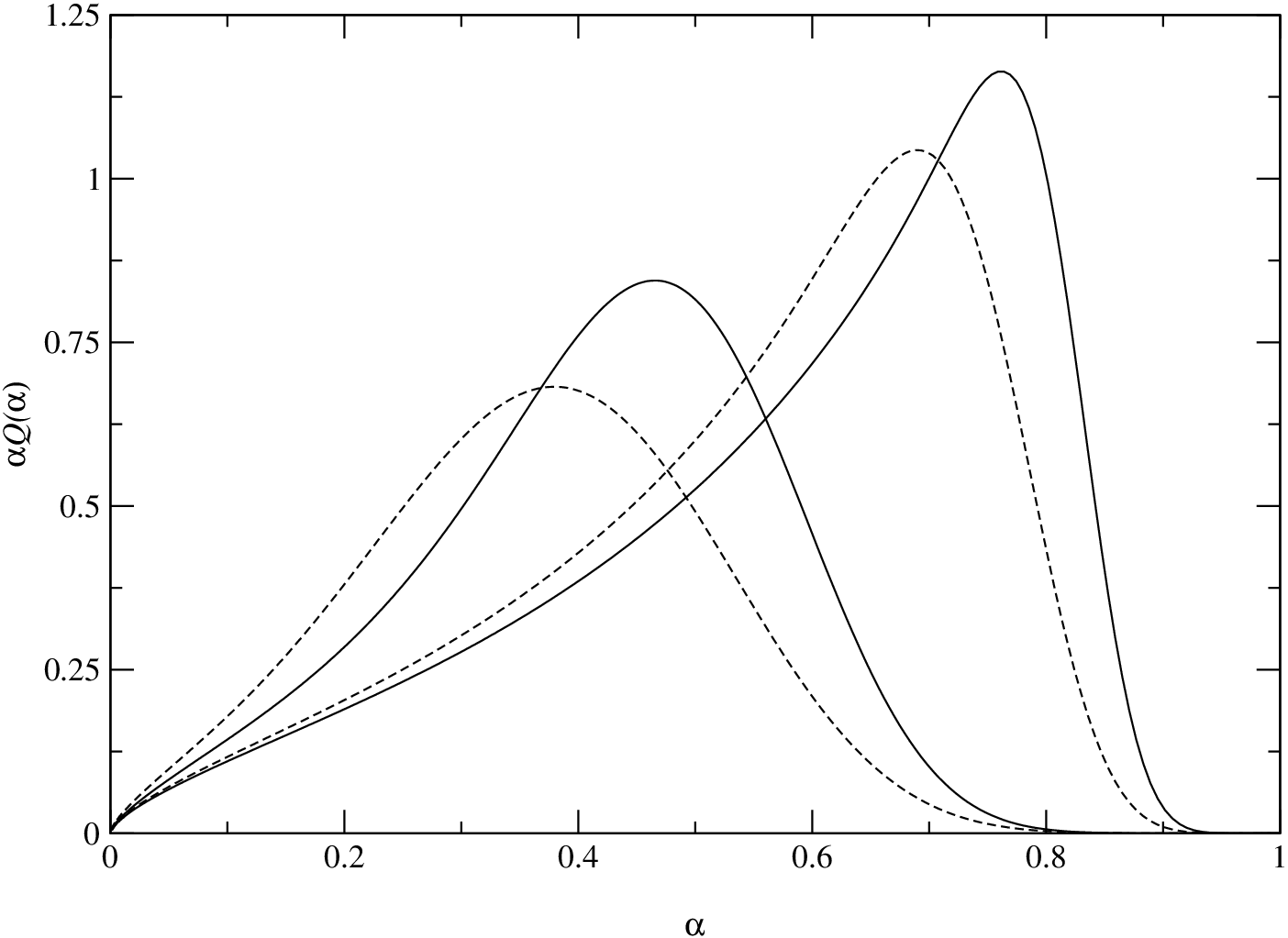}}
\end{center}
\vspace{0.3cm}
\centerline{Fig.6(b)}
\end{figure}

\end{document}